\documentclass{aa}  

\usepackage{booktabs}

\usepackage{graphicx}
\usepackage{natbib}
\usepackage{txfonts}
\usepackage{threeparttable}
\usepackage{hyperref}
\usepackage{siunitx}
\usepackage{color}
\usepackage{xcolor}
\definecolor{mhi}{rgb}{0.6,0,0.6}

\definecolor{hj}{rgb}{0.4,0.7,0.4} 
\definecolor{msp}{rgb}{0.0,0.6,0.6}

\begin{document}

   \title{The properties of dwarf spheroidal galaxies in the Cen\,A group:}
   \subtitle{Stellar populations, internal dynamics, and a heart-shaped H$\alpha$ ring\thanks{Based on observations collected at the European Organisation for Astronomical Research in the Southern Hemisphere under ESO program 0101.A-0193(A) and 0101.A-0193(B).}\fnmsep \thanks{The fully reduced data cubes are available in electronic form
at the CDS via anonymous ftp to cdsarc.u-strasbg.fr (130.79.128.5)
or via http://cdsweb.u-strasbg.fr/cgi-bin/qcat?J/A+A/.}}
   %Spectroscopic follow-up of dwarf galaxies around Cen\,A with MUSE: evidence for a co-moving plane-of-satellites
\titlerunning{The properties of dwarf spheroidal galaxies in Cen\,A galaxy group}
%A spectroscopic study of dwarf galaxies around Cen\,A with MUSE}
   %\title{E pur si muove: coherent motion of Cen\,A's plane-of-satellites}
   \author{
          Oliver M\"uller\inst{1}
                              \and
          Katja Fahrion\inst{2}
         \and
         Marina Rejkuba\inst{2}
         \and
         Michael Hilker\inst{2}
         \and
         Federico Lelli\inst{3}
         \and
         Katharina Lutz\inst{1}
         \and
         Marcel S. Pawlowski\inst{4}
         \and
         Lodovico Coccato\inst{2}
         \and
         Gagandeep S. Anand\inst{5}
         \and
         Helmut Jerjen\inst{6}
          }
          
           \institute{Observatoire Astronomique de Strasbourg  (ObAS),
Universite de Strasbourg - CNRS, UMR 7550 Strasbourg, France\\
 \email{oliver.muller@astro.unistra.fr}
\and
 European Southern Observatory, Karl-Schwarzschild Strasse 2, 85748, Garching, Germany
\and
School of Physics and Astronomy, Cardiff University, Queens Buildings, The Parade, Cardiff,CF24 3AA, UK
\and
Leibniz-Institut f\"ur Astrophysik Potsdam (AIP), An der Sternwarte 16, D-14482 Potsdam, Germany
\and
Institute for Astronomy, University of Hawaii, 2680 Woodlawn Drive, Honolulu, HI 96822, USA
\and
 Research School of Astronomy and Astrophysics, Australian National University, Canberra, ACT 2611, Australia
}

   \date{Received September 15, 1996; accepted March 16, 1997}

   \abstract{Dwarf spheroidal galaxies (dSphs) have been extensively investigated in the Local Group, 
   but their low luminosity and surface brightness make similar work in {more distant}
   galaxy groups challenging. Modern instrumentation unlocks the possibility of scrutinizing these faint systems in other environments, expanding the parameter space of group properties.
   We use MUSE spectroscopy to study the properties of 
   14 known or suspected dSph satellites of Cen\, A. Twelve targets are confirmed to be group members based on their radial velocities. Two targets are background galaxies at $\sim$50~Mpc: KK 198 is a face-on spiral galaxy, and dw1315-45 is an ultra-diffuse galaxy with an effective radius of $\sim$2300~pc. The 12 confirmed dSph members of the Cen\, A group have 
   old and metal-poor stellar populations and follow the stellar metallicity-luminosity relation defined by the dwarf galaxies in the Local Group. 
   In the three brightest dwarf galaxies (KK\,197, KKs\,55, and KKs\,58), we identify globular clusters, as well as a planetary nebula in KK\,197, although its association with this galaxy and/or the extended halo of Cen A is uncertain.
   Using four discrete tracers, we measure the velocity dispersion and dynamical mass of KK\,197. This dSph appears dark matter dominated and lies on the radial acceleration relation of star-forming galaxies within the uncertainties. {It also is consistent with predictions stemming from modified Newtonian dynamics (MOND).} Surprisingly, in the dwarf KK\,203 we find an extended H$\alpha$ ring. Careful examination of {\it Hubble Space Telescope} photometry reveals a very low level of star formation at ages between 30-300 Myr. The H$\alpha$ emission is most likely linked to a $\sim$40~Myr old supernova remnant, although other 
   possibilities for its origin cannot be entirely ruled out.
   }

   \maketitle
%
%-------------------------------------------------------------------

\section{Introduction}
Dwarf galaxies make up the bulk of galaxies in the Universe \citep{1990A&A...228...42B,1994A&ARv...6...67F}. 
Typically, they are defined as galaxies {less luminous} than {$-17$\,mag in the $V$-band \citep{1994ESOC...49....3T, 2009ARA&A..47..371T}, as galaxies with stellar masses below a few times $10^{9}$ M$_\odot$ \citep{2017ARA&A..55..343B, 2017ApJ...851...22M}, or as galaxies with circular velocities below 100 km s$^{-1}$ \citep{2014A&A...563A..27L}.}
{These {contraints} roughly coincide with the properties of the Small Magellanic Cloud (SMC)} and set dwarf galaxies apart from giant galaxies.
{Dwarf galaxies are} separated into gas-rich dwarfs {with ongoing star formation} -- including dwarf irregulars (dIrrs) and blue compact dwarfs (BCDs) -- and gas-poor dwarfs {with predominantly old stellar populations. The former ones are typically found in the field environment, in galaxy groups, and in the outer parts of galaxy clusters, while the latter are almost exclusively found in galaxy clusters and as satellites of massive spirals and ellipticals \citep{1990A&A...228...42B, 2012ApJ...757...85G}.}

{For historical reasons, the taxonomy of gas-poor dwarfs is complex \citep{1994ESOC...49...13B} and includes three common nomenclatures: (i) Dwarf ellipticals (dEs) have been predominantly found in galaxy clusters \citep{1984AJ.....89..919S} and {make up} the bright end of the dwarf galaxy population from $\sim$10$^9$ down to $\sim$10$^7$ $L_\odot$; (ii) dwarf spheroidals (dSphs) have traditionally been studied in the Local Group \citep{1998ARA&A..36..435M} and constitute an intermediate luminosity {range} from $\sim$10$^{7}$ to $\sim$10$^{5}$ $L_\odot$; and (iii) ultra-faint dwarfs (UFDs) were discovered in the Local Group after the advent of the Sloan Digital Sky Survey (SDSS, \citealt{2000AJ....120.1579Y}) and other deep optical surveys and represent the {faintest}
%dimmest 
galaxies known, with luminosities from $\sim$10$^5$ to $\sim$10$^3$ $L_\odot$ \citep[e.g.,][]{2005ApJ...626L..85W,2007ApJ...654..897B,2015ApJ...805..130K,2015ApJ...808L..39K,2020ApJ...890..136M}. All these gas-poor dwarfs, however, form a single sequence in structural diagrams comparing stellar luminosity, effective radius, and effective surface brightness \citep{2009ARA&A..47..371T, 2009ApJS..182..216K}. In fact, next-generation surveys of the Virgo and Fornax clusters have started to probe gas-poor dwarfs down to luminosities of $\sim$10$^{5} L_\sun$ \citep{2016ApJ...824...10F, 2018ApJ...855..142E, 2019A&A...625A.143V}, blurring the traditional difference between cluster dEs and Local Group dSphs.}

The most detailed studies of {dSphs} {have been carried out in} the Local Group  \citep[e.g.,][]{2004ApJ...617L.119T,2009ARA&A..47..371T,2006AJ....131..895K,2006A&A...459..423B,2011MNRAS.411.1013B,2008ApJ...684.1075M,2010A&A...524A..58T,2011ApJ...727...78K,2013ApJ...778..103H,2014ApJ...789..147W,2018A&A...618A.122T,2020A&A...635A.152T}, {where we can observe their stars down to faint main sequence evolutionary phases and measure individual stellar
{velocities} and their chemical abundances, thereby constraining the formation and evolution of the hosts. Due to their low luminosities and low surface brightnesses, the physical characterization of dSphs beyond the Local Group is {inherently difficult}. However, as these low-mass systems hold some of the fundamental observational constraints for cosmology \citep{2017ARA&A..55..343B}, there is a growing effort {to search for and study them}
in a range of environments, {as}
enabled by modern, highly sensitive instruments} 
\citep[e.g.,][]{2001A&A...371..487J,2006A&A...448..983R,2008AJ....135..380L,2010ApJ...708L.121D,2010A&A...521A..43L,2010A&A...516A..85C,2011A&A...530A..58C,2011ApJ...739....5W,2013MNRAS.428.2980R,2015A&A...581A..82M,2015ApJ...799..172T,refId0,2019ApJ...874L..12D,2019A&A...629L...2M,2019ApJ...884...79C, 2019ApJ...885..153B,2019A&A...625A..94H,2019A&A...625A.143V,2020MNRAS.tmp.2144S,2020MNRAS.495.2582G}.

{Globular clusters (GCs) do exist in all major galaxies. In the regime of dwarf galaxies, however, the {presence} of GC systems around them {becomes} stochastic \citep[e.g., ][]{2010MNRAS.406.1967G}.
In the Milky Way system}, only {four} dwarf galaxies {have a population of} old and massive globular clusters. These are the SMC \citep{2003MNRAS.338..120M}, the {tidally} disrupting Sagittarius dwarf spheroidal \citep{2003MNRAS.340..175M}, Fornax \citep{2003MNRAS.340..175M}, and the ultra-faint dwarf Eridanus 2 \citep{2016ApJ...824L..14C}. The remaining dwarfs host no known GCs \citep{2020arXiv200514014H}. {Thus, a key question arises:} Are the Milky Way dwarf galaxies a representative sample of the general population of dwarf galaxies \citep[see e.g.,][]{2017ApJ...847....4G}? In galaxy clusters, a plethora of GCs have been found associated with dwarf galaxies \citep[e.g., ][]{2006AJ....132.2333S,2019MNRAS.484.4865P,2020MNRAS.492.4874F}, 
with {observed numbers of GCs larger than} for the Milky Way dwarfs. This is explained by a {higher GC formation efficiency in dwarf galaxies living} in denser environments  \citep{2008ApJ...681..197P}. However, {for} the intermediate range of galactic environments there is still a lack of data. Surveys like the Mass Assembly of early Type gaLAxies with their fine Structures (MATLAS, \citealt{2015MNRAS.446..120D}) or the Dwarf Galaxy Survey with Amateur Telescopes (DGSAT, \citealt{2016A&A...588A..89J}) aim to {fill this gap by} targeting massive elliptical  and spiral galaxies in the nearby field ($10<D<45$\,Mpc). 

The closest giant elliptical galaxy {to us} -- {Centaurus\,A (Cen\,A)} at a distance of 3.8\,Mpc \citep{2010PASA...27..457H} -- is the ideal environment for {detailed} studies of dwarf galaxies since it is possible to investigate their physical properties
{using both} deep imaging observations that resolve individual bright stars \citep{2002A&A...385...21K,2006A&A...448..983R,2010A&A...516A..85C,2012A&A...541A.131C,2019ApJ...872...80C,MuellerTRGB2019} and imaging of 
%from 
their integrated light properties \citep{2016ApJ...823...19C,2016arXiv160807285T,2017MNRAS.469.3444T,2018ApJ...867L..15T,2017A&A...597A...7M}. {Regarding spectroscopic 
observations, the} most recent integral-field observations of the 
two dSphs KKs58 and KK197 in the Cen\,A group revealed nuclear star clusters (NSCs) and GCs associated with them \citep{2020A&A...634A..53F}. The simultaneous extraction of the metallicity of {both} the NSC and 
the {galaxy} stellar body showed that {these} NSCs are more metal-poor {than their hosts}, suggesting
that they have formed through the {process of} in-spiraling GCs. Intriguingly, the masses, sizes, and metallicities of {these} two NSCs are consistent with the properties of  known ultra-compact dwarfs \citep{2011MNRAS.414.3699M,2018ApJ...858...20V,2020arXiv200102243V,2019A&A...625A..50F}, suggesting that stripped nuclei of dwarf galaxies are the progenitors of low-mass ultra-compact dwarfs \citep{2013ApJ...775L...6S}. 

These two dwarf galaxies {are part of} a larger survey aiming to study the phase-space distribution of the dwarf galaxy satellites around Cen\,A \citep{2015ApJ...802L..25T,2016A&A...595A.119M,2018Sci...359..534M}. {The survey has two {main components: (i) deep imaging to resolve individual bright red giant stars and measure distances to dwarf galaxies by means of the tip of the red giant branch (TRGB) method}, and (ii) integrated light spectroscopy to measure {the dwarfs'} radial velocities. In the first part we used FORS2@VLT to {derive accurate distances, thereby also confirming group membership}, and measure mean photometric metallicities and structural parameters for nine dwarf satellites of Cen\,A \citep{MuellerTRGB2019}.} In this article, we present the {MUSE spectroscopic} data analysis of 14 putative dwarf galaxies in our survey, study their individual properties, and compare them the to Local Group dwarf galaxies. 
In a companion paper, we will explore the dynamical properties of the whole satellite system.

\section{Observations and data reduction}
\label{observations}
The data {were} acquired with the MUSE integral field spectrograph mounted at UT4 of the VLT {on Cerro}
Paranal, Chile \citep{Bacon2010} as part of a 46 hour program (PI: M\"uller, proposal ID: 0101.A-0193) {designed to measure distances {(requiring 26h with FORS2)} and  }{line-of-sight} velocities {(20h with MUSE}) of dwarf galaxy candidates.
The targets were selected from \citet{2017A&A...597A...7M} and  \citet{2004AJ....127.2031K,2013AJ....145..101K} as
{the most likely members of the Cen\,A group that could be detected by MUSE with reasonable integration times.}
We used the Wide Field Mode (WFM) of MUSE, {providing} a field-of-view (FOV) of 1\arcmin$\times$1\arcmin\, sampled at $0\farcs2$~pix$^{-1}$. The observed {galaxies} have sizes of $\sim$1$\arcmin$, which fit well in the MUSE FOV. {The instrument was used in its nominal wavelength setting covering 480-930 nm with a mean resolution of 3000\footnote{The resolving power ranges between 1770 at 480 nm and 3590 at 930nm in the WFM.}.}
Given that our {primary} goal was to obtain {the systemic}
velocity by binning the signal from the entire target galaxy, {it was possible to conduct} the observations 
under relatively poor seeing conditions and sometimes with thin clouds, meaning filler conditions,
in service mode between April and June 2018. %{These observations have revealed that 12 out of our 14 targets are actual members of the Cen\,A group. The remaining two targets are background star-forming galaxies at larger distances (see below).}

Depending on the {galaxy's} {mean} surface brightness we required one or two Observation Blocks (OBs). For {eight} targets with higher surface brightness, the single OB included four science exposures ($O$), each 500\,s long, interleaved with two offset sky exposures ($S$) of 250\,s. Hence, the OB had the following sequence of exposures: {\it OSO\,OSO}. The other six targets with 
{lower} surface brightness had two OBs per target, each with an {\it OSO} strategy. In this case, the science 
and offset exposures had 1160\,s and 580\,s, {respectively.}
We employed {our} FORS2 images -- where possible -- to select empty regions for the offset sky exposures, otherwise we used DECam or DSS images. 

\begin{figure*}[ht]
    \centering
    \includegraphics[width=0.24\linewidth]{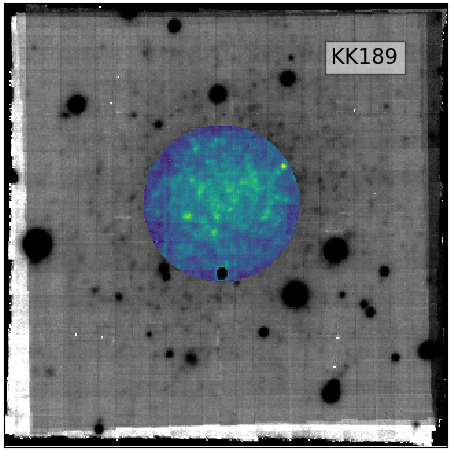}
    \includegraphics[width=0.24\linewidth]{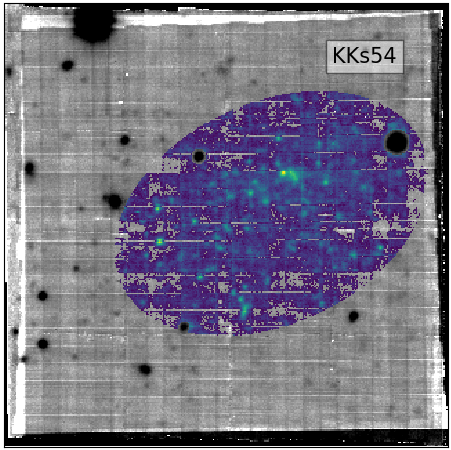}
    \includegraphics[width=0.24\linewidth]{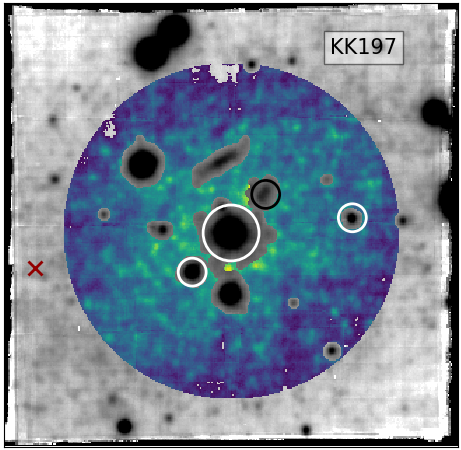}
    \includegraphics[width=0.24\linewidth]{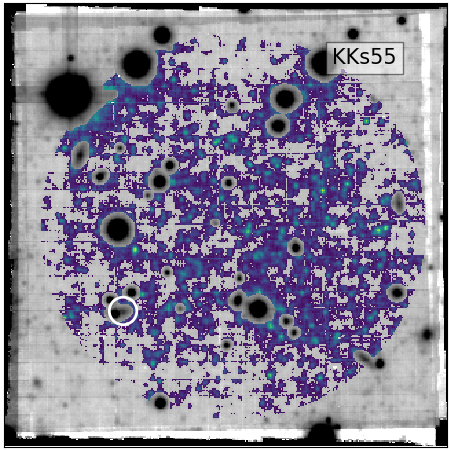}\\
    \includegraphics[width=0.24\linewidth]{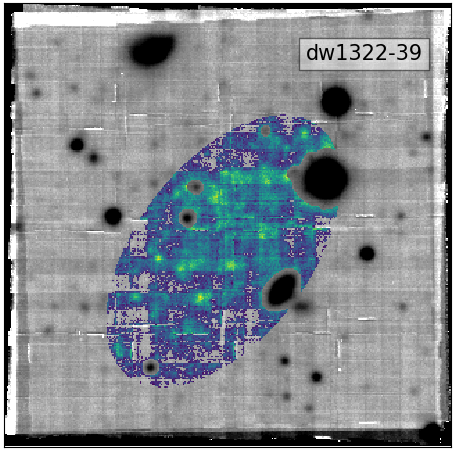}
    \includegraphics[width=0.24\linewidth]{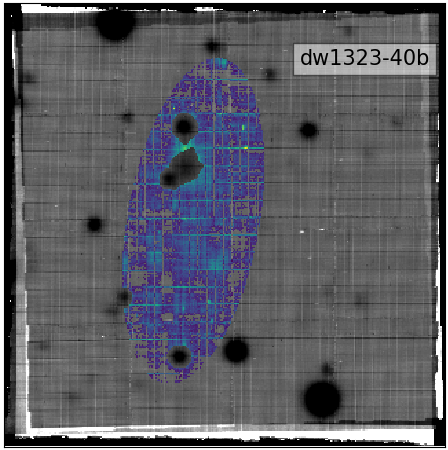}
    \includegraphics[width=0.24\linewidth]{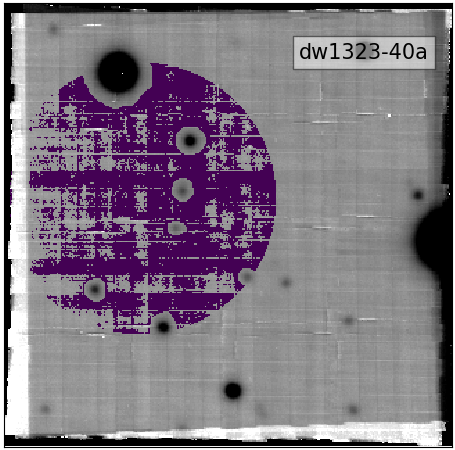}
    \includegraphics[width=0.24\linewidth]{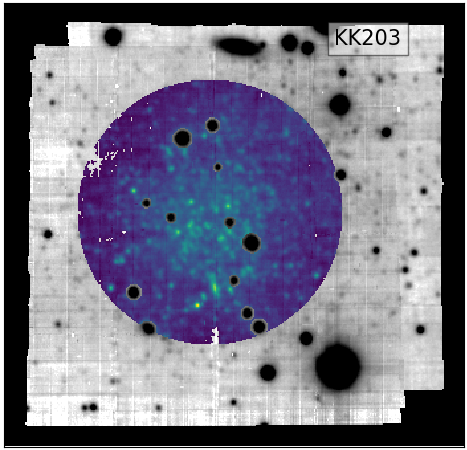}\\
    \includegraphics[width=0.24\linewidth]{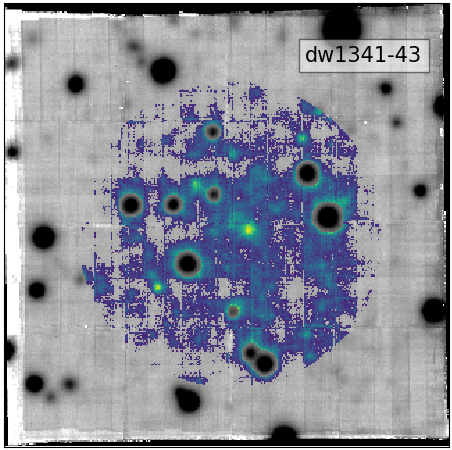}
    \includegraphics[width=0.24\linewidth]{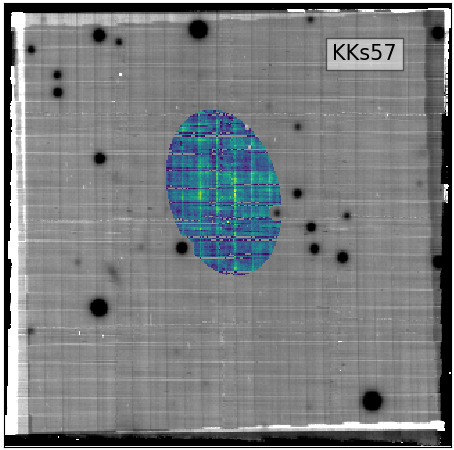}
    \includegraphics[width=0.24\linewidth]{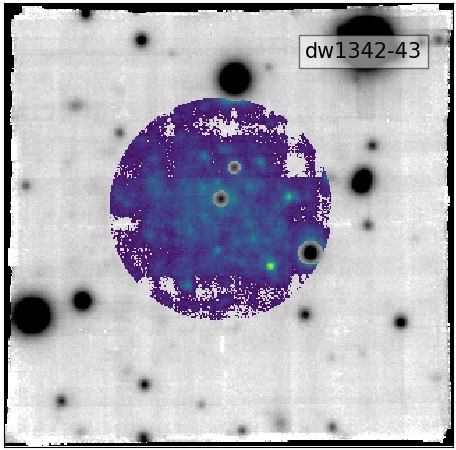}
    \includegraphics[width=0.24\linewidth]{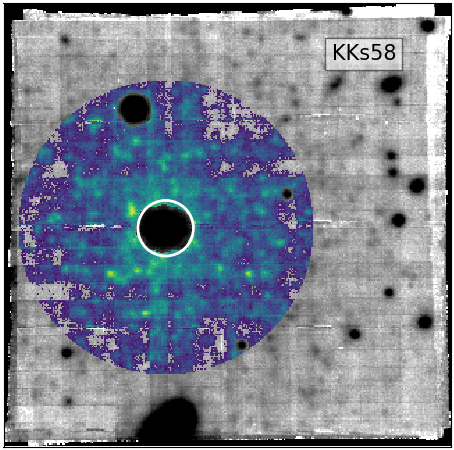}\\
    \caption{{Collapsed, white-light images} 
    of the dwarf galaxies {confirmed to be Cen\, A group members}.  The regions where we extracted the spectra are indicated in color. {The NSC are indicated with the large white circles, the GCs as small white circles, the stellar association as small black circle, and the position of the PN is marked with a red cross. The images show the entire MUSE FOV -- 1\arcmin per side -- which corresponds to 1.1\,kpc at a mean distance of the Cen\,A group (3.8\,Mpc),  and are oriented with north up and east to the left.} %The {Nucleated Star Clusters in the centre} of KK\,197 and KKs58 are marked with a small cross.
    }
    \label{fig:ifu}
\end{figure*}

The MUSE Internal Data Products are available from the ESO Science Archive, which includes products run through the MUSE pipeline version 2.2 \citep{Weilbacher2012,2020arXiv200608638W}. The data have been preprocessed, bias and flat-field corrected, astrometrically calibrated, sky-subtracted, wavelength and flux calibrated \citep{Hanuschik2017}\footnote{see also \url{http://www.eso.org/observing/dfo/quality/PHOENIX/MUSE/processing.html}}. 
The sky subtraction {makes use of the} offset sky exposures.
To further reduce the sky residual lines, we applied the Zurich Atmosphere Purge (ZAP) principal component analysis algorithm \citep{Soto2016}. We tested whether the offset sky exposure or empty patches on the science exposure lead to better results in removing sky residuals. For objects having at least 20\% of sky in the science exposure, we got better results {with these empty patches} than using the offset sky exposure, therefore we used the former where applicable. To select the empty sky patches on the science exposures we ran 
%source detection using
{the MTObjects tool} \citep{teeninga2013bi,teeninga2015improved}, which creates a segmented fits file with all {detected} 
%astronomical 
sources. This segmentation map {served as input} mask for ZAP. In some cases, we also manually masked the outskirts of the target galaxies or undetected objects to further improve the quality of the sky subtraction.

\begin{table*}[ht]
\small
\caption{Observation summary for the 14 dwarf galaxy candidates. 
}% title of Table
     % is used to refer this table in the text
\centering                          % used for centering table
\begin{tabular}{l l l r r r l l l l}        % centered columns (4 columns)
\hline\hline                 % inserts double horizontal lines
Name & $\alpha_{2000}$ & $\delta_{2000}$ & Observing& Nexp  & Exptime  & AM & IQ & S/N & Note\\    % table heading 
     &(hh:mm:ss.s)       & (dd:mm:ss)      & Date     &  &   (s)  & & (") & &\\    % table heading 
& (1) & (2) & (3)  & (4)  & (5) & (6) & (7) & (8) & (9)\\ 
\hline      \\[-2mm]                  % inserts single horizontal line
%\\
{KK189}& 13:12:45.1 & 
                      {$-$41:49:55} & 7/8 May 2018 & 4 & 500 & 1.07 & 0.7 & 28.5 & thin clouds\\
{KKs54}& 13:21:32.3 & $-$31:53:10 & 7/8 May 2018 & 4 &500 & 1.09 & 0.7 & 12.1 & thin clouds\\
{KK197}& 13:22:01.9 & $-$42:32:07 & 15/16 Apr 2018 & 2 & 1160 & 1.20 & 1.2 & 27.6 & clear sky\\
&  &  & 8/9 May 2018 & 3 &1160 & 1.05 & 1.2 & & extra exposure due to thin/thick clouds\\
{KKs55}& 13:22:12.4 & $-$42:43:50  & 19/20 May 2018 & 2 &1160 & 1.05 & 0.8 & 14.4 & clear/thin\\
&  &  & 20/21 May 2018 & 2 & 1160 & 1.09 & 0.7 &  &thin clouds\\
{dw1322-39}& 13:22:32.0 & $-$39:54:19  &  	8/9 May 2018 & 4 & 500 &  1.17& 1.5 & 11.9 &thin clouds\\
{dw1323-40b}& 13:23:55.0 & $-$40:50:08 &  	5/6 Jun 2018 & 4 & 1160 & 1.11 & 1.3 & 13.1 &thin/clear\\
{dw1323-40a}& 13:24:53.0 & $-$40:45:40 & 8/9 May 2018 & 4 &500 & 1.35 & 1.7 & 9.3 & thin clouds\\
{KK203}& 13:27:27.6 & $-$45:21:08  &  	19/20 May 2018 & 5 & 1160 & 1.28 & 0.6 & 24.0 & extra exposure due to a technical issue\\
{dw1341-43}& 13:41:36.9 & $-$43:51:16 &  	5/6 Jun 2018 & 4 & 1160 & 1.06 & 1.3 & 12.9 &mostly clear sky\\
{KKs57}& 13:41:37.9  & $-$42:34:54 & 7/8 May 2018 & 4 & 500 & 1.20 & 0.8 & 13.3 & thin clouds\\
{dw1342-43}& 13:42:43.9 & $-$43:15:18 &  	11/12 May 2018 &4 &500 & 1.16 & 1.4 & 17.8 &clear sky\\
{KKs58}& 13:46:00.1 & $-$36:19:43  &  	8/9 May 2018 &4 &500 & 1.04 & 1.5 & 17.7 & thin clouds\\
{KK198}& 13:22:56.0 & $-$33:34:21  &  	9/10 Jun 2018 & 4 &500 & 1.09 & 0.8 & --- & not Cen\,A group member; thin clouds\\
{dw1315-45}& 13:15:56.0 & $-$45:45:02 & 7/8 May 2018 & 4 & 1160 & 1.28 & 0.6 & --- & not Cen\,A group member; thin clouds\\
%{name}& RA & $-$DEC & DATE & $4\times500 $ & X & $0.0-0.0$ & SNR\\

\hline
\end{tabular}
\tablefoot{(1) and (2): Coordinates of the center of observation (epoch J2000); (3) Date of observation; 
(4 \& 5): {Number of exposures on target and {integration} time per exposure};
(6) Average airmass during observation; and (7): Average image quality (seeing).
(8): The measured signal-to-noise ratio on the stacked cube {between 6600 and 6800\,\si{\angstrom} (see text)}. }
\label{fit} 
\end{table*}

We extracted the integrated spectra {using} 
an elliptical aperture, which was derived from the optical images of the galaxies. Foreground stars and background objects were masked using the python implementation \textit{sep} \citep{SEP} of Source Extractor \citep{SExtractor}. 
Furthermore, we masked pixels with negative mean flux as a simple constrain on the signal-to-noise ratio (S/N) per spaxel.
The size of the aperture was chosen to optimize the S/N. 
Fig.\,\ref{fig:ifu} {shows} the white light images with highlighted target extraction areas.

\subsection{Radial velocity and stellar population measurements}

\begin{figure*}[ht]
    \centering
    \includegraphics[width=\linewidth]{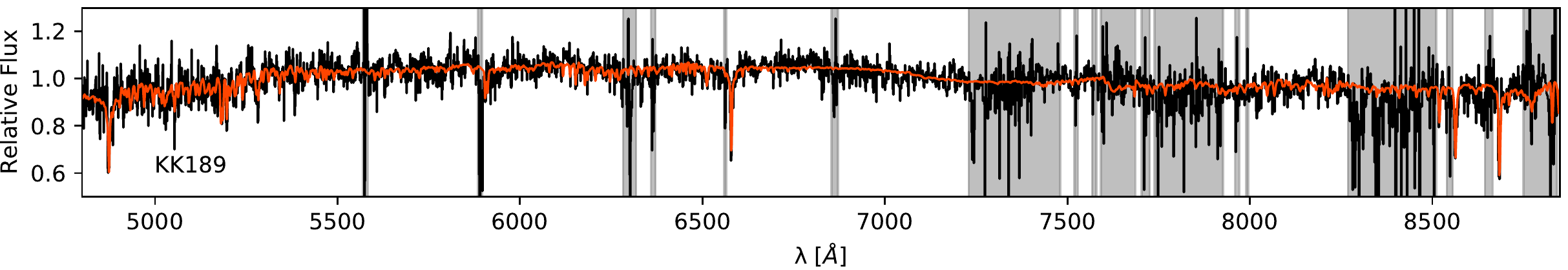}\\
    \includegraphics[width=\linewidth]{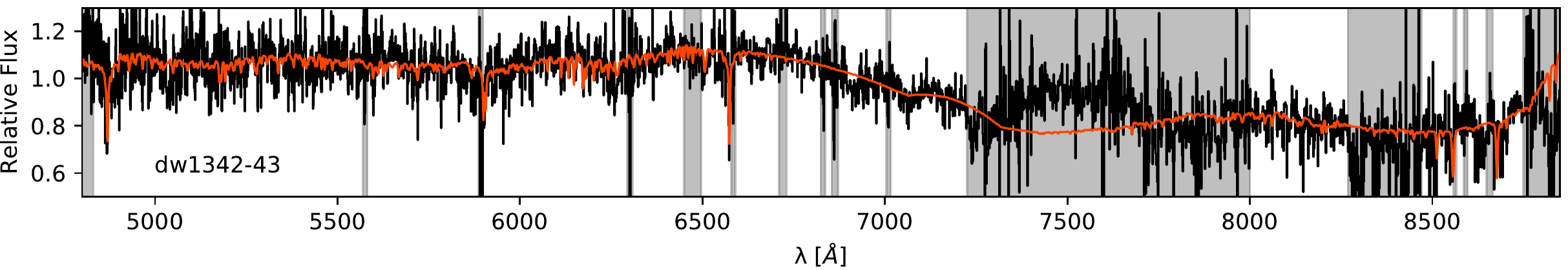}\\
    \includegraphics[width=\linewidth]{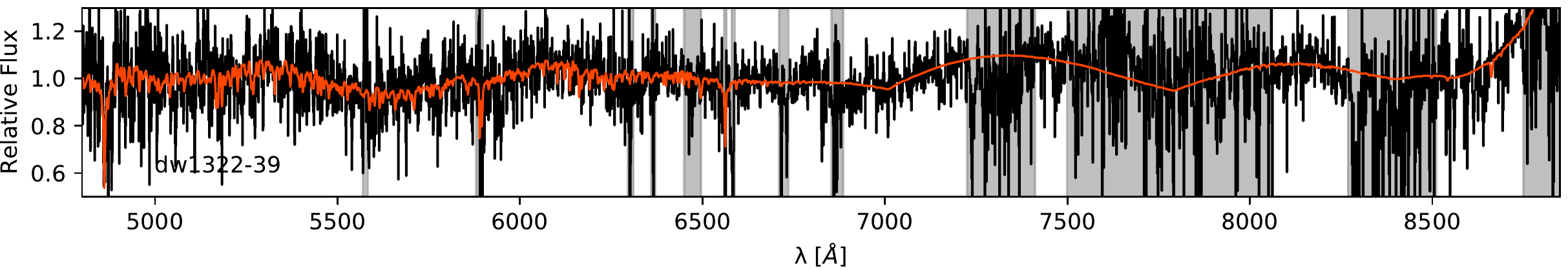}\\
    \caption{{Integrated MUSE} spectra (black) of three {example} dwarf galaxies from our sample. The S/N ratio decreases from top (S/N=28.5) to bottom (S/N=11.9) ({see Table \ref{fit}}). The gray area are masked regions, the red lines correspond to the best-fit from pPXF. The spectra of the remaining galaxies are shown in Figs.\,\ref{fig:spectra_all1} and \ref{fig:spectra_all2} in the Appendix.}
    \label{fig:spectra}
\end{figure*}

To extract line-of-sight velocities and stellar population properties from the galaxy spectra, we employed pPXF \citep{2004PASP..116..138C}, a standard routine in fitting line-of-sight velocity distributions from absorption-line spectra. We followed the same methodology as described in several previous studies \citep{2019A&A...625A..76E,2019A&A...625A..77F,2020A&A...640A.106M}. 
In short, we use a set of Single Stellar Population (SSP) spectra from the eMILES library \citep{2016MNRAS.463.3409V}, with metallicities {[M/H]} ranging from solar down to $-$2.27 dex and ages from 70 Myr to 14.0 Gyr. {We assume} a Kroupa initial mass function (IMF, \citealt{2001MNRAS.322..231K}).
The spectra from the SSP library are convolved with the line-spread function as described in \citet[][see also the Appendix of \citealt{2019A&A...625A..76E}]{2017A&A...608A...5G}. A variance spectrum was measured on the masked data cube and added to pPXF to improve the fitting. 
For the {kinematic} fit, we used 8 {and 12} degrees of freedom {for} the multiplicative and additive polynomials, respectively \citep{2019A&A...625A..76E}. {For the age and metallicity fits, we fixed the velocity and used no additive polynomials, but kept the 12th degree in the multiplicative polynomial \citep{2019A&A...625A..77F}. Then we used the weights provided by pPXF to derive the mean metallicities and mean ages from the SSP models for each galaxy. Similarly, we calculated the stellar mass-to-light ratio from the weights given by pPXF and the photometric predictions from the eMILES library \citep{2010MNRAS.404.1639V} in the $V$-band.} 
Fig.\,\ref{fig:spectra} shows the spectra and the best pPXF fits for three dwarfs. To improve the fits, we masked the remaining sky lines, which were not removed by ZAP. The errors {on the best-fit parameters} were estimated
{with a Monte Carlo method} where we reshuffle the residuals in a bootstrap approach. The 16 and 84 percent interval (1$\sigma$ in frequentist statistics) give the uncertainty. 
The S/N ratio per pixel is measured in a region between 6600 and 6800\,\si{\angstrom} {devoid of strong absorption or emission lines}. It is calculated as the mean fraction between the flux and the square root of the variance. The variance itself has to be multiplied with the $\chi^2$ value estimated by pPXF.

A note on the velocity extraction of KK\,197 and KKs\,58. In \citet{2020A&A...634A..53F} we derived for these two galaxies line-of-sight velocities of 643.2$\pm$3.5\,km/s and 482.6$\pm$12.6\,km/s, respectively. This differs slightly from the values measured here, due to a different set of SSP models used, as well as different ways of extracting the spectra from the cube. However, the values here and the values in \citet{2020A&A...634A..53F} are consistent within 1$\sigma$. The same goes for the the metallicities and ages. In \citet{2020A&A...634A..53F} we estimated a slightly more metal-rich ($-$0.84$\pm$0.12\,dex and $-$1.35$\pm$0.23\,dex) and younger (10$\pm$1.0\,Gyr and $\sim$7\,Gyr) population than we estimate here. For the metallicity, this is within 3$\sigma$ and 1$\sigma$, respectively. The age estimation from data with a low S/N ratio is highly uncertain, therefore a discrepancy within a few Gyr is not uncommon.

\subsection{Globular clusters and planetary nebulae detection}

To search for any GCs, we {identified} all the point sources with Source Extractor and extracted their spectra. For this purpose, we created a 2D image by collapsing the cube along the wavelength axis. We %have 
{then} inspected all the individual spectra to create a catalog of GCs. Two of our galaxies -- KKs\,58 and KK\,197 -- have already been analyzed in \citet{2020A&A...634A..53F}. {These two dSphs} contain a NSC, and KK\,197 furthermore hosts two GCs and an extended stellar association. Here, we find one new GC in KKs\,55, which is also visible in Hubble Space Telescope (HST) images available for this galaxy. For the remaining dwarfs, the data is either too shallow or {there are} simply 
%contain 
no GCs {within the MUSE FOV}.

{Furthermore, we searched the MUSE cubes for planetary nebulae (PNe)
that might serve as additional kinematic tracers. For this purpose, we derived a narrow band image for each galaxy at the expected redshifted position of the [OIII]$\lambda$5007 \AA\ emission line. By subtracting this {narrow-band image} from a {collapsed} image of the {nearby stellar} continuum, PNe should stand out as residual point sources. 
{Using this approach we {detected one PN} in KK\,197} { at the following location: 13:22:04.51/$-$42:32:14.09, which is 0.5\,arcmin from the center of the galaxy}.
Its spectrum is shown in Fig.\,\ref{fig:emission_pne} {indicating the location of typical emission lines shifted to the velocity of the object. Only the strongest emission lines are visible due to too low S/N}. The other galaxies either contain no PNe within the MUSE FOV, or the data is too shallow to detect them.} 

\section{Results}
In this section, we discuss the 
{overall} properties of the {stellar populations of the} observed dwarf galaxies {(Sect. \ref{Sec:stPop}). {Our radial velocity measurements confirm the Cen\,A group membership for 12 out of 14 targets.} Then,}
we look into their GC and PNe population {(Sect. \ref{Sec:GC} and \ref{Sec:PNe}) and derive the velocity dispersion and dynamical mass of KK\,197 (Sect. \ref{Sec:KK197}) using these discrete tracers. We also report the surprising discovery of a heart-shaped extended H$\alpha$ ring in the dSph KK\,203 (Sect. \ref{Sec:KK203}).}

\subsection{Properties of the stellar populations}\label{Sec:stPop}

\begin{table*}[ht]
\caption{Derived properties of the dwarf galaxies.
}% title of Table     % is used to refer this table in the text
\centering                          % used for centering table
\tiny
\setlength{\tabcolsep}{2pt} % https://tex.stackexchange.com/questions/16519/adding-space-between-columns-in-a-table
\begin{tabular}{l c c c c c c c c c c c c}        % centered columns (4 columns)
\hline\hline                 % inserts double horizontal lines
 & KK\,189 & KKs\,54 & KK\,197 & KKs\,55 &dw1322-39 & dw1323-40b & dw1323-40a & KK\,203 & dw1341-43 & KKs\,57& dw1342-43 & KKs58  \\    % table heading 
\hline      \\[-2mm]    % inserts single horizontal line 
$v$ (km s$^{-1}$) & 
$ 752.6$$\pm$$4.0$ &  %KK\,189  changed
$ 621.3$$\pm$$10.6$ &% KKs\,54
$ 642.7$$\pm$$2.9$  &% KK\,197
$ 550.0$$\pm$$23.7$&% KKs\,55 changed
$ 656.3$$\pm$$9.7$&% dw1322-39
$ 497.0$$\pm$$12.4$ &% dw1323-40b
$ 450.0$$\pm$$14.2$ &% dw1323-40a
$ 305.9$$\pm$$9.5$ &% KK\,203 changed
 $ 636.4$$\pm$$14.1$ &% dw1341-43
$ 511.3$$\pm$$16.8$ &% KKs57
 $ 510.3$$\pm$$8.1$ &% dw1342-43
$ 476.5$$\pm$$5.2$ % KKs58 
\\ \addlinespace[0.05cm] 
[Fe/H] (dex) &
$ -1.43^{+0.07}_{-0.07}$ & %KK\,189
 $ -1.81^{+0.07}_{-0.26}$ &% KKs\,54
$ -1.15^{+0.12}_{-0.01}$ &% KK\,197
$ -1.14^{+0.04}_{-0.30}$ &% KKs\,55
$ -1.79^{+0.22}_{-0.13}$ &% dw1322-39
 $ -1.84^{+0.01}_{-0.32}$&% dw1323-40b
$ -1.95^{+0.30}_{-0.14}$&% dw1323-40a
$ -1.75^{+0.11}_{-0.28}$ &% KK\,203
$ -1.79^{+0.03}_{-0.33}$ &% dw1341-43
 $ -1.90^{+0.07}_{-0.27}$ &% KKs57
$ -1.69^{+0.13}_{-0.19}$ &% dw1342-43
$ -1.49^{+0.07}_{-0.09}$ % KKs58 
\\ \addlinespace[0.05cm] 
age (Gyr) &
$ 7.6^{+1.2}_{-1.2}$ & %KK\,189
$ [7.4, 11.2] $&% KKs\,54
$ 11.7^{+2.2}_{-0.1}$ &% KK\,197
$ 5.8^{+3.3}_{-0.5}$&% KKs\,55
$ 11.5^{+2.5}_{-0.7}$ &% dw1322-39
$ [13.8, 14.0] $ &% dw1323-40b
$ 12.2^{+1.8}_{-3.7}$ &% dw1323-40a
$ [7.0, 12.4] $ &% KK\,203
$ [8.3, 11.7] $ &% dw1341-43
$ [10.7, 14.0] $ &% KKs57
$ 12.2^{+1.8}_{-1.8}$ &% dw1342-43
$ 14.0^{+0.0}_{-2.3}$% KKs58 
\\ \addlinespace[0.05cm] 
M$_{\rm V}$/L$_{\rm V}$ &
$ 1.4^{+0.1}_{-0.1}$ & %KK\,189
 $ [1.3, 1.8] $ &% KKs\,54
$ 2.1^{+0.5}_{-0.0}$&% KK\,197
$ 1.2^{+0.4}_{-0.1}$&% KKs\,55
$ 1.8^{+0.3}_{-0.1}$&% dw1322-39
 $ [2.1, 2.2] $ &% dw1323-40b
$ 1.9^{+0.3}_{-0.5}$ &% dw1323-40a
$ 2.1^{+0.0}_{-0.1}$ &% KK\,203
$ [1.4, 1.9] $ &% dw1341-43
 $ [1.6, 2.2] $ &% KKs57
$ 1.9^{+0.2}_{-0.2}$  &% dw1342-43
$ 2.2^{+0.0}_{-0.3}$% KKs58 
\\ \addlinespace[0.05cm] 
\\
M$_{\rm V}$ (mag)&
$-11.2$& %KK\,189
$-10.4$&% KKs\,54
$-13.0$&% KK\,197
$-12.6$&% KKs\,55
$-10.0$&% dw1322-39
$-10.0$&% dw1323-40b
$-10.4$&% dw1323-40a
$-11.7$&% KK\,203
$-10.1$&% dw1341-43
$-10.6$&% KKs57
$-9.8$&% dw1342-43
$-11.9$% KKs58 
\\ \addlinespace[0.05cm] 
$r_{\rm eff}$ (arcsec)&
14.4& %KK\,189
16.6&% KKs\,54
44.4&% KK\,197
36.4&% KKs\,55
20.7&% dw1322-39
17.1&% dw1323-40b
15.2&% dw1323-40a
19.8 &% KK\,203
20.2&% dw1341-43
12.0&% KKs57
15.5&% dw1342-43
26.4% KKs58 
\\ \addlinespace[0.05cm]
$\mu_{\rm  eff, V}$ (mag/$\Box$)&
24.5 & %KK\,189
25.6&% KKs\,54
24.8&% KK\,197
25.5&% KKs\,55
25.9&% dw1322-39
26.1&% dw1323-40b
25.4&% dw1323-40a
24.6 &% KK\,203
26.2&% dw1341-43
24.8&% KKs57
25.5&% dw1342-43
24.6% KKs58 
\\ \addlinespace[0.05cm] 
\hline % ^{+}_{-}
\end{tabular}
\label{properties} 
\tablefoot{The properties for $M_{V}$,  $r_\text{eff}$, and $\mu_{\rm  eff, V}$ were derived using the photometry from \citet{2000AJ....119..593J} for KKs58, and from \citet{2017A&A...597A...7M} for the rest. No values for $\mu_{\rm  eff, V}$ and $r_\text{eff}$ are available in the literature for KK\,203, so we derived it here ourselves (see Appendix \ref{photometry}).}
\end{table*}

{The integrated spectra of the 12 dSphs (see, e.g., Fig.\,\ref{fig:spectra}) display several absorption features and no strong emission lines, as expected for passive galaxies dominated by old stars. The velocities derived from these absorption lines are consistent with the velocity range of the Centaurus group \citep{2018Sci...359..534M}. 
The pPXF fits indicate that these spectra are consistent with old ($6-14$ Gyr) and metal poor { ($-1.95 \lesssim \text{[Fe/H]}\lesssim -1.15$\,dex)} stellar populations.} For an accurate estimation of the {weighted mean ages of the stellar populations, however,} we would need {high} S/N ratios (see Fig.\,1 in the Appendix of \citealt{2019A&A...628A..92F} for the required S/N ratios for {measuring} velocities, metallicities, and ages with E-MILES SSP templates), which is not reached here. Nevertheless, we have estimated ages for all the objects and where pPXF did not converge to a single value, we give the 80\% bounds. 
All the properties of the dwarfs are compiled in Table\,\ref{properties}, {but the mean stellar ages should be taken with a grain of salt}. It is interesting to note that for the dwarf galaxy with {one of} the highest S/N ratio -- KK\,203 -- we find the youngest age ($7.6\pm1.2$\,Gyr). 
%\mhi{[Would it be interesting to note that KK 203, having one of the highest S/N values, shows the youngest age boundary (7 Gyr) of all galaxies? Maybe a hint towards young stellar populations in this galaxy, as revealed from the CMD analysis. Of course, we don't detect the very young ages in the integrated spectrum, but overall this galaxy might have hold on its gas for a longer time, thus showing an overall younger average age as the other galaxies!?]}
For two targets we measure strong emission lines, for which the derived velocities put them far in the background (see Appendix\,\ref{Sec:background}).

\begin{figure}[ht]
    \centering
    \includegraphics[width=9cm]{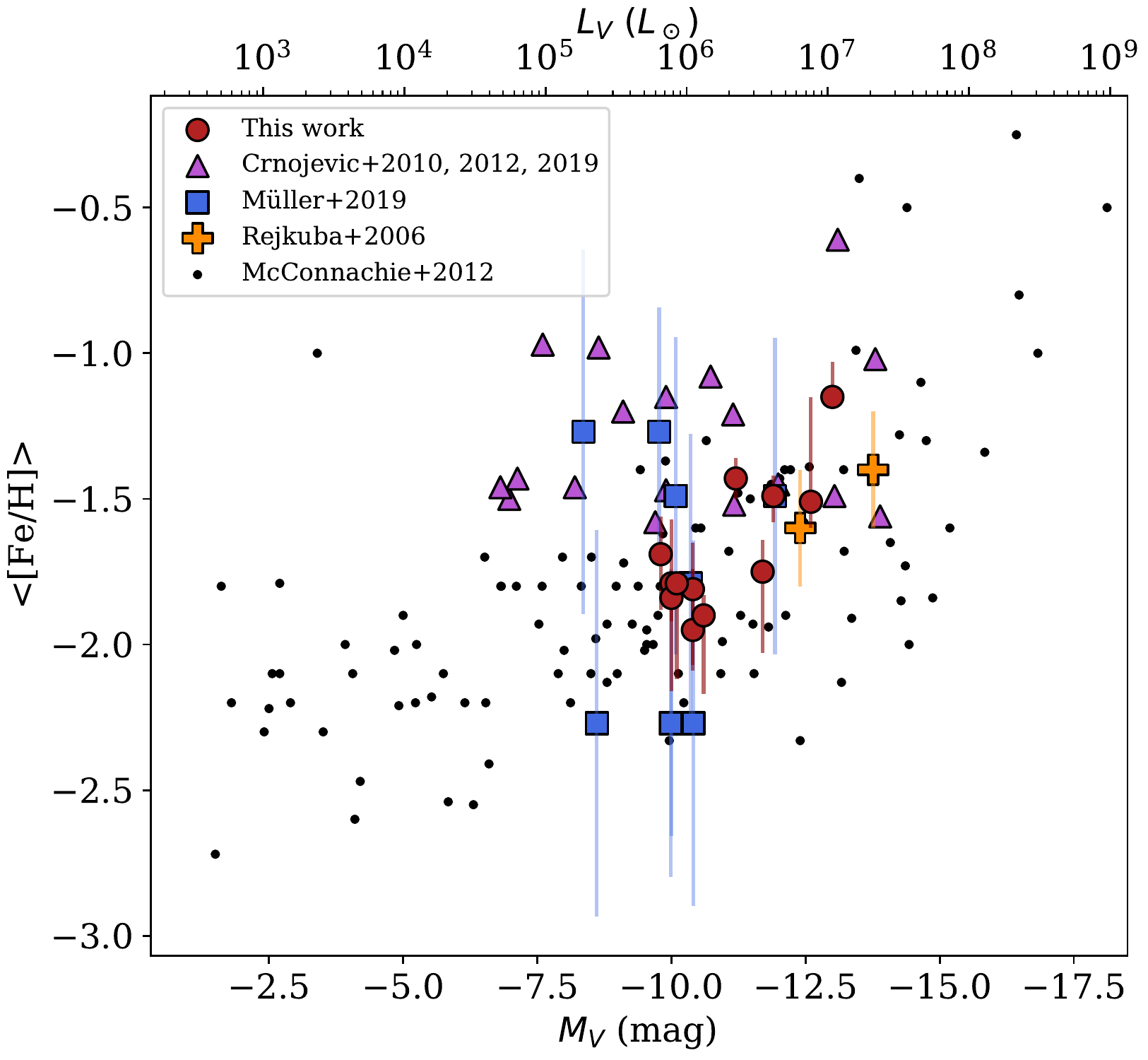}
    \caption{Luminosity-metallicity relation for dwarf galaxies in the Cen\,A group \citep[crosses, triangles, squares][]{2006A&A...448..983R, 2010A&A...516A..85C,2011A&A...530A..59C,2019ApJ...872...80C,MuellerTRGB2019} and Local Group (black dots, \citealt{2012AJ....144....4M}, 
    from the updated online catalog, and are compiled from different sources, for example from \citealt{2004MNRAS.354.1263B,2009ApJ...705..758M,2009MNRAS.397L..26C,2013ApJ...779..102K,2016MNRAS.463..712T,2016ApJ...818...39H,2018ApJ...860...66M}). The dwarfs from this work are indicated as red dots.}
    \label{fig:metallicity}
\end{figure}

Dwarf galaxies in the Local Group follow several scaling relations \citep{2008ApJ...684.1075M,2013ApJ...779..102K}. {One of the most important is} the {stellar} metallicity-luminosity relation. 
In Fig.\,\ref{fig:metallicity} we {investigate this} relation for {our 12 dSphs}
and compare them {with a} {compilation of Local Group and nearby galaxies from \citet{2012AJ....144....4M}, as well as with previously studied dwarfs  in the Cen\,A group} \citep{2006A&A...448..983R,2010A&A...516A..85C,2012A&A...541A.131C,2019ApJ...872...80C,MuellerTRGB2019}.
We note that the {mean} metallicity {for these different samples is based on several} different {techniques}. 
{The extended catalog from \citet{2012AJ....144....4M} includes metallicity measurements based on photometric methods such as RGB colors, isochrones fitting, or full CMD fitting, as well as spectroscopic metallicities from low-resolution Ca triplet or from spectral synthesis based on medium to high-resolution spectra. The latter measurements are based on observations of individual stars and it is interesting therefore to compare alongside with the mean metallicity of Cen\,A dwarfs obtained from resolved stellar population studies \citep{2006A&A...448..983R,2010A&A...516A..85C,2012A&A...541A.131C,2019ApJ...872...80C,MuellerTRGB2019}, as well as integrated light analysis with MUSE.} 
The dwarf galaxies of the Cen\,A group follow {the same stellar} metallicity-luminosity relation as dwarf galaxies in the Local Group ({Fig.~\ref{fig:metallicity}). While the MUSE spectroscopic metallicities lie exactly on top of the \citet{2012AJ....144....4M} metallicity-luminosity relation, there is a much larger scatter for photometric measurements. This is not surprising because at the distance of the Cen\,A group, only few magnitudes of the upper RGB can be resolved, and from optical photometry it is impossible to distinguish between 5 and 10~Gyr old RGB stars. The age-metallicity degeneracy and possibly incorrect assumptions for the uniform old ($\sim 10$~Gyr) age of the population lead to larger uncertainties in photometric metallicities. Furthermore, a systematic offset in metallicity scale may result from a choice of a specific isochrone set or empirical RGB calibration \citep{2014ApJ...791L...2R,2014A&A...563A...5S}.}

\subsection{Globular clusters properties}\label{Sec:GC}

\begin{figure}[ht]
    \centering
    \includegraphics[width=9cm]{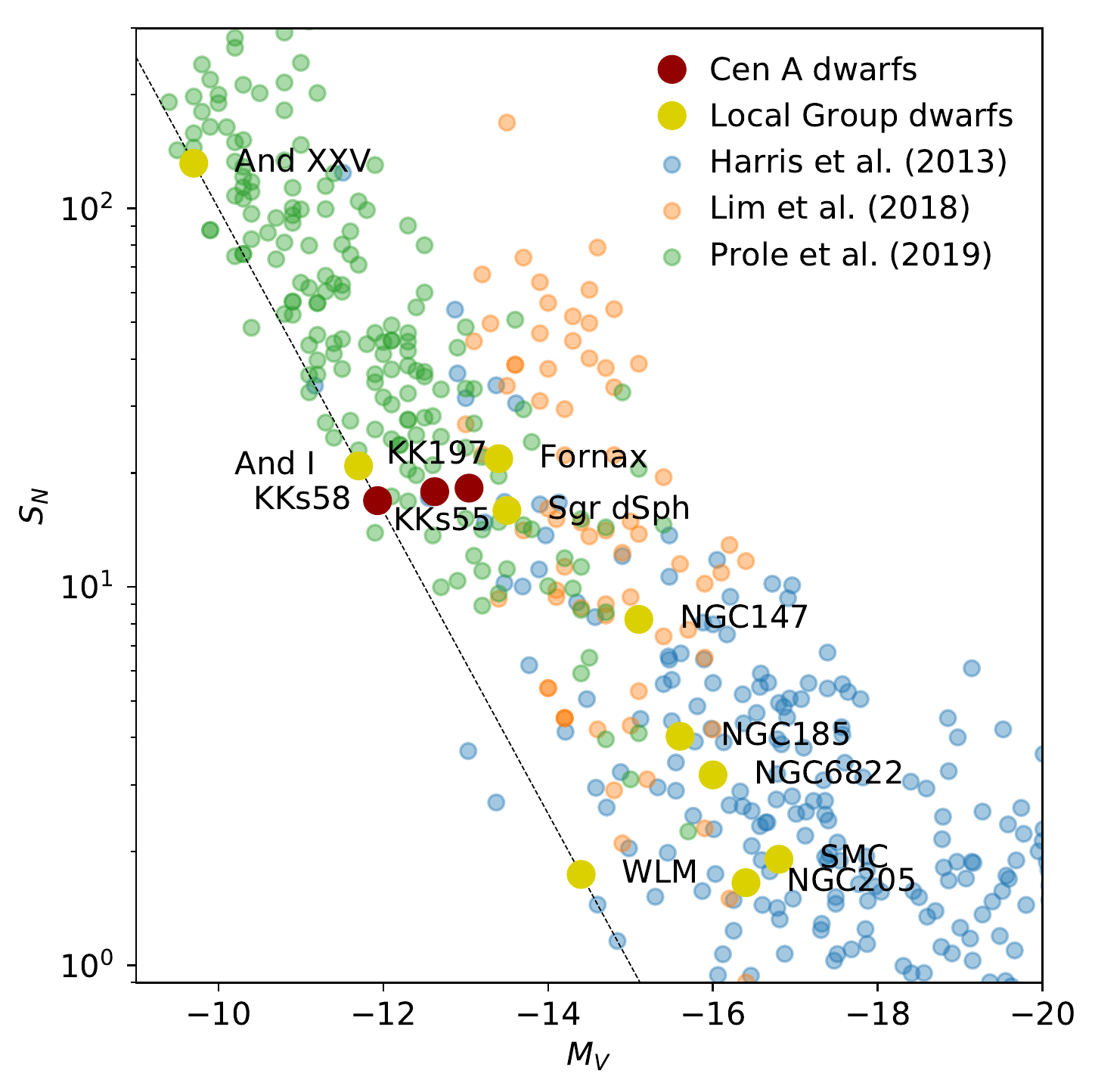}
    \caption{Specific frequency of the Cen\,A dwarfs (big red dots) compared to other galaxies. Blue dots are nearby galaxies from \citet{2013ApJ...772...82H}, orange dots are Coma dwarf galaxies \citep{2018ApJ...862...82L}, and green dots are Fornax dwarf galaxies \citep{2019MNRAS.484.4865P}. {The Local Group dwarfs are indicated with big yellow dots \citep{2016IAUS..312..157G,2016ApJ...829...26C,2017PASA...34...39C}. The dashed line indicates the specific frequency of an object with one GC.}
    {The specific frequencies for the  Coma and Fornax dwarfs were estimated on a statistical basis, and often they are consistent with zero within the uncertainties.}
    }
    \label{fig:specific}
\end{figure}

In our sample of 12 dwarf galaxies, three host stellar clusters: KK\,197, KKs\,55, and KKs\,58. \citet{2020A&A...634A..53F} analyzed two GCs, one NSC, and one extended stellar association in KK\,197 and one NSC in KK\,58. Around KKs\,55, \citet{2010MNRAS.406.1967G} found one GC based on deep HST data {that lies outside the MUSE FOV}. For this galaxy, we discovered an additional GC {in the MUSE data} that is visible in the HST data as well, although it can easily be confused with a background galaxy shining through {KKs\,55}. Its {spectrum} {unambiguously} confirms the association with the dwarf galaxy. {The coordinates for this new GC are RA=13:22:13.869 and Dec=$-$42:44:05.04. Performing simple aperture photometry on the HST data we derive apparent magnitudes of $m_V=22.8\pm0.1$\,mag and $m_I=21.8\pm0.1$\,mag in the Vega system.  With a distance modulus of 27.93\,mag \citep{2013AJ....145..101K}, this gives extinction corrected absolute magnitudes of $M_V=-5.5$\,mag and $M_I=-6.2$\,mag, respectively, with a color of $(V-I)_0=0.7$\,mag.}
{From the MUSE spectrum}, we determined a velocity of  $531.4\pm15.4$\,km\,s$^{-1}$, a metallicity of  $-1.50^{+0.34}_{-0.07}$\,dex, a {mean} age of $ 12.9^{+1.4}_{-2.5}$\,Gyr, and a mass-to-light ratio of $ 2.1^{+0.3}_{-0.4}$ M$_\odot$/L$_\odot$. This GC seems to be more metal-poor than the stellar population of its host dwarf galaxy ([M/H]$= -1.14^{+0.04}_{-0.30}$\,dex).

To characterize the GC systems, we derived the specific frequency $S_N = N_{\text{GC}} \times 10^{0.4(M_V + 15)}$ \citep{Harris1981} for each dwarf. {The results are presented} in Fig\,\ref{fig:specific}. Where no GC is detected, $S_N$ is {assigned a value of zero, {although} it is possible that some GCs are outside the MUSE FOV or are too faint to be detected in our data}. For KK\,197 and KKs\,58, the  $S_N$ are 18.2 and 16.9, respectively \citep{2020A&A...634A..53F}, and for KKs\,55 it is 17.9. These numbers are compatible  with the specific frequencies { of dwarf galaxies from the Local Group \citep{2016IAUS..312..157G}} and other nearby dwarf galaxies \citep{2010MNRAS.406.1967G,2013ApJ...772...82H}, as well as dwarf galaxies in clusters \citep{2018ApJ...862...82L,2019MNRAS.484.4865P,2020arXiv200614630S}, {but we note that the $S_N$ has a significant scatter in the classical dwarf galaxy regime.}

\subsection{Planetary nebula properties}\label{Sec:PNe}
Do we expect to find PNe in our dwarf galaxies? {Considering} the {small} number of PNe in Local Group dwarfs, we expect {very few} PNe in our sample \citep{2015IAUGA..2249670R}. {The number of PNe scales with the total sampled luminosity as $N_{\mathrm{PN}} = \alpha L_{\mathrm{bol}}$ \citep{2006MNRAS.368..877B}. There is a considerable uncertainty in theoretical predictions for the value of $\alpha$, which depends on the PN lifetime and specific evolutionary flux (i.e., how many PNe are produced per simple stellar population luminosity), which in turn depends on the star formation rate and metallicity of the population. The Local Group PN census confirms theoretical expectations that metal-poor populations host fewer PNe. For example for the SMC metallicity  of $\sim -1.25$~dex \citet{2015IAUGA..2249670R} derived $\alpha = 1 \mathrm{PN} / 4.6 \times 10^6$ L$_\odot$. 
Some of our dwarfs are significantly more metal-poor than the SMC and similar to Local Group dSph that have no known PN. Hence, it is not surprising that we only detect one} PN, which is {potentially} associated with the brightest dwarf in our sample: KK\,197. 

\begin{figure}[ht]
    \centering
    \includegraphics[width=9cm]{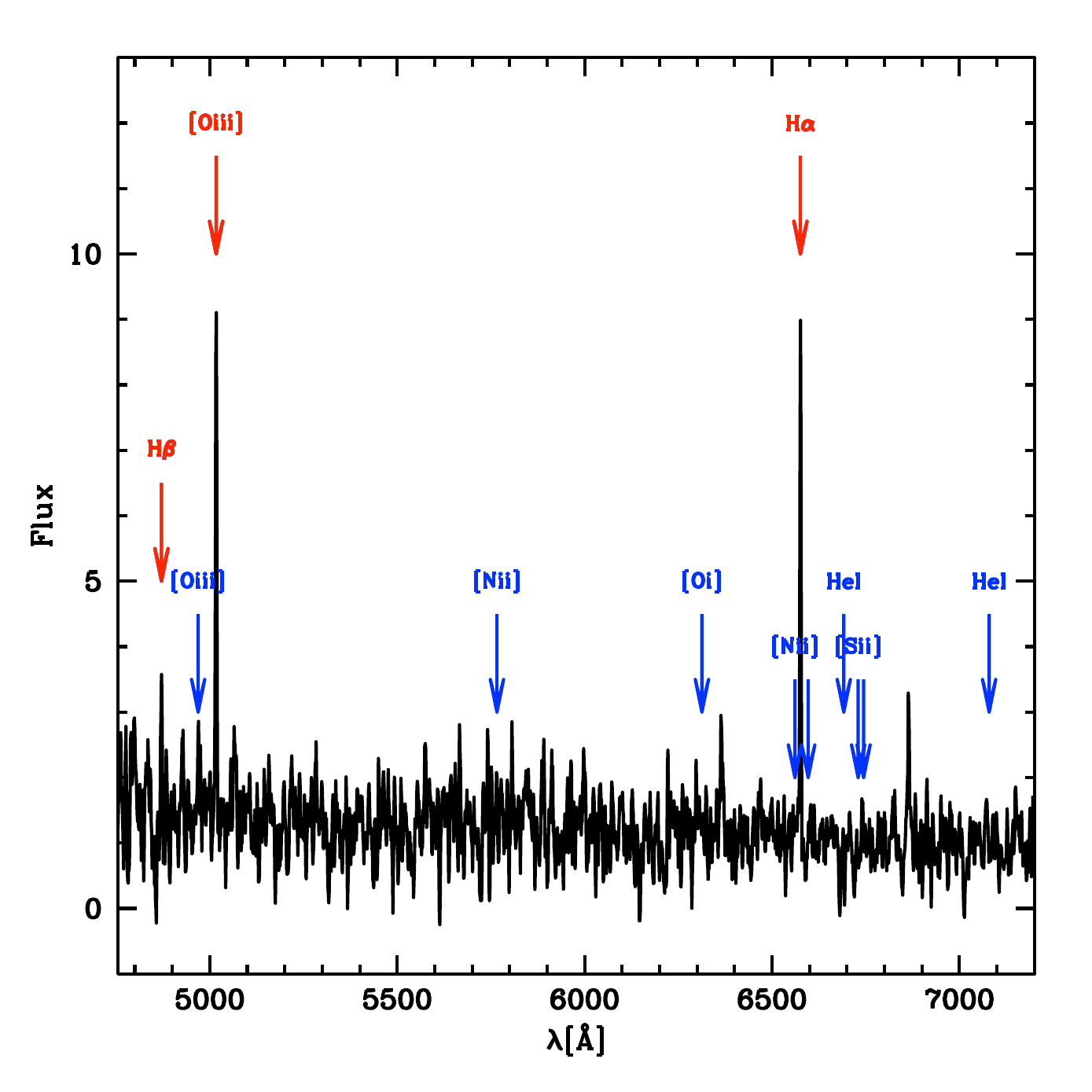}
    \caption{{MUSE spectrum of the PN in KK\,197}. The red {and blue} arrows indicate typical emission lines of PNe {of which only the strongest ones ($H\alpha$, 5007\AA\, [OIII] and possibly $H\beta$ are detected, marked in red}. {The spectra were slightly smoothed with a three-point boxcar filter.}}
    \label{fig:emission_pne}
\end{figure}

In Fig.\,\ref{fig:emission_pne} we present the spectrum for {this} PN and indicate the typical emission lines {of PNe}. There is clear [OIII] and H$\alpha$ emission and likely H$\beta$ emission. {The line-of-sight velocity of the PN from the emission lines ($v_{\rm KK197, PN}=588.42 \pm 2.5$ km s$^{-1}$) differs by about 55 km s$^{-1}$ from the systemic velocity of KK\,197 {measured from the stellar absorption lines in the same datacube} ($642.7 \pm 2.9$ km s$^{-1}$), which may be interpreted as a real physical association.}
However, it is also possible that the PN is associated with Cen\,A itself and appears in the MUSE FOV only by chance, as the PNe population of the Cen\,A halo extends to the position of KK\,197 \citep{2015A&A...574A.109W}. With an offset of $\sim$50 km s$^{-1}$ to the mean of the PNe population ($v_{\text{Cen\,A}}= 536.7\pm4.2$\,km/s) it is entirely plausible that the PN is an interloper, since the standard deviation of the 1107 PNe studied in \citet{2015A&A...574A.109W} is 141\,km s$^{-1}$.  

{\subsection{Internal dynamics of KK\,197}}
\label{Sec:KK197}
\begin{figure*}[ht]
    \centering
    \includegraphics[width=0.45\linewidth]{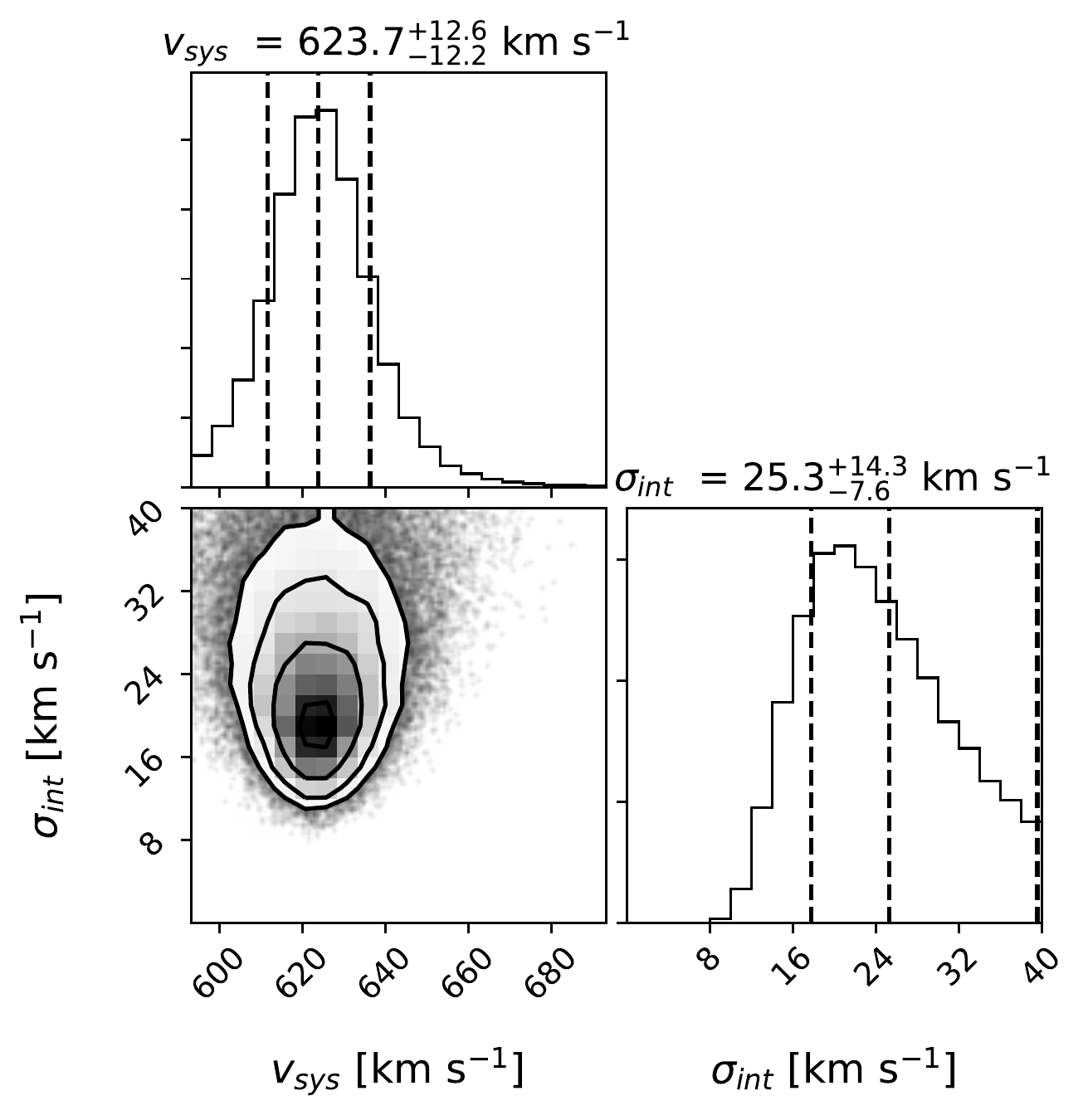}
    \includegraphics[width=0.45\linewidth]{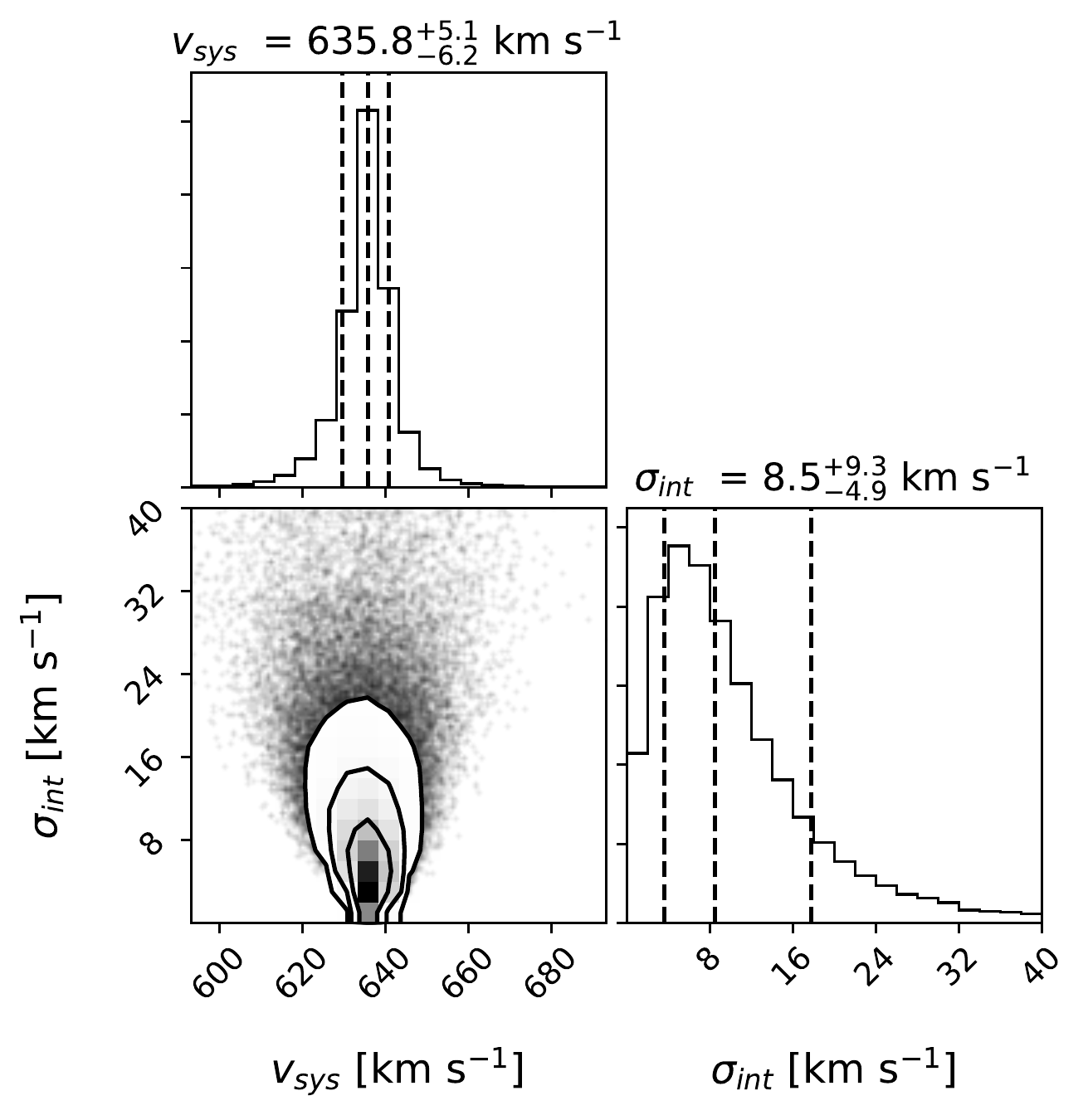}
 \caption{Posterior distribution from the MCMC analysis of the velocity dispersion of KK\,197, as well as the systemic velocity of the tracers. Left is the estimation including the PN as tracer, right without including it. The three dashed lines indicate the 16, 50, and 84 percentiles, which correspond to the upper and lower uncertainty boundaries, and the best parameter estimation (i.e., the median).}
    \label{fig:mcmc}
\end{figure*}

The simultaneous extraction of the spectrum of the stellar body {of a galaxy} and the spectra of {its associated} star clusters 
makes it possible to estimate the {galaxy's velocity dispersion\footnote{{To directly measure the velocity dispersion of the stellar body, we would need deeper data. See for example \citet{2019A&A...625A..76E} for a study with MUSE of the velocity dispersion measured on the spectra of a comparable dwarf galaxy.}} and therefore} probe {its dynamical mass and} dark matter content \citep[e.g.,][]{2018Natur.555..629V,2019ApJ...874L...5V,2018ApJ...859L...5M,2020A&A...640A.106M}. More and more such studies are now {conducted in the}
dwarf galaxy regime. A reliable dynamical mass estimate, {however,} requires the presence of multiple kinematic tracers of the underlying {gravitational} potential, such as GCs or PNe \citep{Cote2001, Pota2013, Forbes2017,2019MNRAS.484..245L,Fahrion2020b}. {For the brightest galaxy in our sample -- KK\,197 --} five 
{discrete} tracers are available. 
For the detailed methodology, we refer to \citet{2020A&A...640A.106M}. The tracers are listed in Table \ref{gc_tracers}.

\begin{table}[ht]
\caption{Kinematic tracers for the mass derivation of KK\,197. The velocities are from \citet{2020A&A...634A..53F} for the stellar clusters and from here for the PN.
}% title of Table     % is used to refer this table in the text
\centering                          % used for centering table
\begin{tabular}{l c c }        % centered columns (4 columns)
\hline\hline                 % inserts double horizontal lines
Name & $v$ & $v_{err}$\\
 & (km/s) & (km/s) \\
 \hline
KK197 & 642.7 &2.9  \\ 
\\
KK197-NSC & 635.4 &1.5  \\
KK197-1 & 636.4 &16\\
KK197-3 & 642.6 &3.8 \\
KK197-SA & 619.3 &10.3\\
KK197-PN & 588.4 &2.5 \\
\hline % ^{+}_{-}
\end{tabular}
\label{gc_tracers} 
\end{table}

In short, we used a Markov Chain Monte Carlo (MCMC) approach with the logarithmic likelihood function given as
\begin{equation}
\log \mathcal{L}=\sum_{i=1}^N{\log \left(\frac{1}{\sqrt{2\pi} \sigma_{\rm obs}}\right) -\frac{(v_{{\rm obs},i}-v_{\rm sys})^2}{2\sigma_{\rm obs}^2}},
\end{equation}
where $N$ is the number of tracers, $v_{\text{obs}, i}$ is the observed {line-of-sight} velocity {of an individual tracer $i$}, $v_{\rm sys}$ is the systemic velocity of {the whole system}, 
and $\sigma_{\rm obs}$ is the observed velocity dispersion given by a combination of the true velocity dispersion $\sigma_{\rm int}$ and the observational uncertainties:
\begin{equation}
\sigma_{\rm obs}^2 =\sigma_{\rm int}^2+\delta_{v,i}^2,
\end{equation}
The two variables $v_{\rm sys}$ and $\sigma_{\rm int}$ are the parameters we are interested in. A non-informative prior is imposed to suppress a too small velocity dispersion (Agnello \& Bruun in prep.):
 \begin{equation} %p=sigma_int/(sigma_int**2+np.mean(v_err)**2)**(3/2)
 P(\theta)=\frac{\sigma_{\rm int}}{(\sigma_{\rm int}^2+\epsilon^2)^{3/2}},
\end{equation}
where $\epsilon$ is the mean velocity error.
 We used 100 walkers with a chain length of 10000 each. The sampled posterior distribution is shown in Fig.\,\ref{fig:mcmc} {for two different cases: considering the newly discovered PN bound to KK\,197 (left, with $N=5$) or assuming it is an interloper associated with Cen\,A (right, with $N=4$).}

{From a dynamical perspective, it is very unlikely that the PN belongs to KK\,197 because the line-of-sight velocity difference of $\sim$55 km s$^{-1}$ to the putative host is comparable to the inferred escape velocity $V_{\rm esc}\simeq\sqrt{2}\sqrt{3}\sigma_{\rm int}$ when the PN is included in the estimate of $\sigma_{\rm int}$ ($V_{\rm esc}\simeq62$ km s$^{-1}$) and much larger when the PN is excluded ($V_{\rm esc}\simeq21$ km s$^{-1}$). Moreover, the systemic velocity of KK 197 from the stellar absorption lines ($643.2 \pm 3.5$ km s$^{-1}$) is closer to the maximum-likelihood value of $v_{\rm sys}$ when the PN is excluded ($635.8^{^+5.1}_{-6.2}$ km s$^{-1}$) rather than when it is included ($623.7^{^+12.6}_{-12.2}$ km s$^{-1}$). In the following, for the sake of completeness, we provide dynamical mass measurements using both values of $\sigma_{\rm int}$ but the one excluding the PN should be considered more reliable despite the lower number of kinematic tracers.}

From the {intrinsic}
velocity dispersion, the dynamical mass enclosed within the de-projected half light radius $r_{1/2}$ can be estimated 
{using the following formula}
\citep{2010MNRAS.406.1220W}:
\begin{equation}
M_{\rm dyn}(r_{1/2})=\frac{4r_{\rm eff}\sigma_{\rm int}^2}{G}
\end{equation}
with $r_{1/2}=4/3 r_{\rm eff}$. The {resulting} dynamical mass within $r_{1/2}$ is {$M_\text{dyn} =0.5^{+ 1.8 }_{-0.4}\times10^{8}$\,M$_\odot$ considering only the GCs and} $4.9^{+ 7.1}_{-2.5}\times10^{8}$\,M$_\odot$ {including the uncertain PN}. To estimate the dynamical mass-to-light ratio, we transform the $V$-band magnitude of KK\,197 into solar luminosities. This gives $M_{\rm dyn}/L_V=4.0^{+ 13.8 }_{-3.3}$\,M$_\odot$/L$_\odot$ and $36.6^{+ 53.2}_{-18.7}$\,M$_\odot$/L$_\odot$ {excluding and including the PN, respectively}.

{The dynamical mass-to-light ratios of galaxies are known to anti-correlate with galaxy luminosity and surface brightness \citep[e.g.,][]{1998ApJ...499...41M}. A more comprehensive representation of galaxy dynamics is offered by the radial acceleration relation (RAR, \citealt{2017ApJ...836..152L}), in which the observed kinematic acceleration $g_{\rm obs}$ is compared to the Newtonian gravitational field $g_{\rm bar}$ from the baryonic mass distribution.} If the data follows the line of unity, {the observed dynamics can be fully explained by the visible matter.}
If the observed acceleration is larger than the one caused by the baryons, an additional {gravitational} component (i.e., dark matter), is necessary. 

\begin{figure}[ht]
    \centering
    \includegraphics[width=9cm]{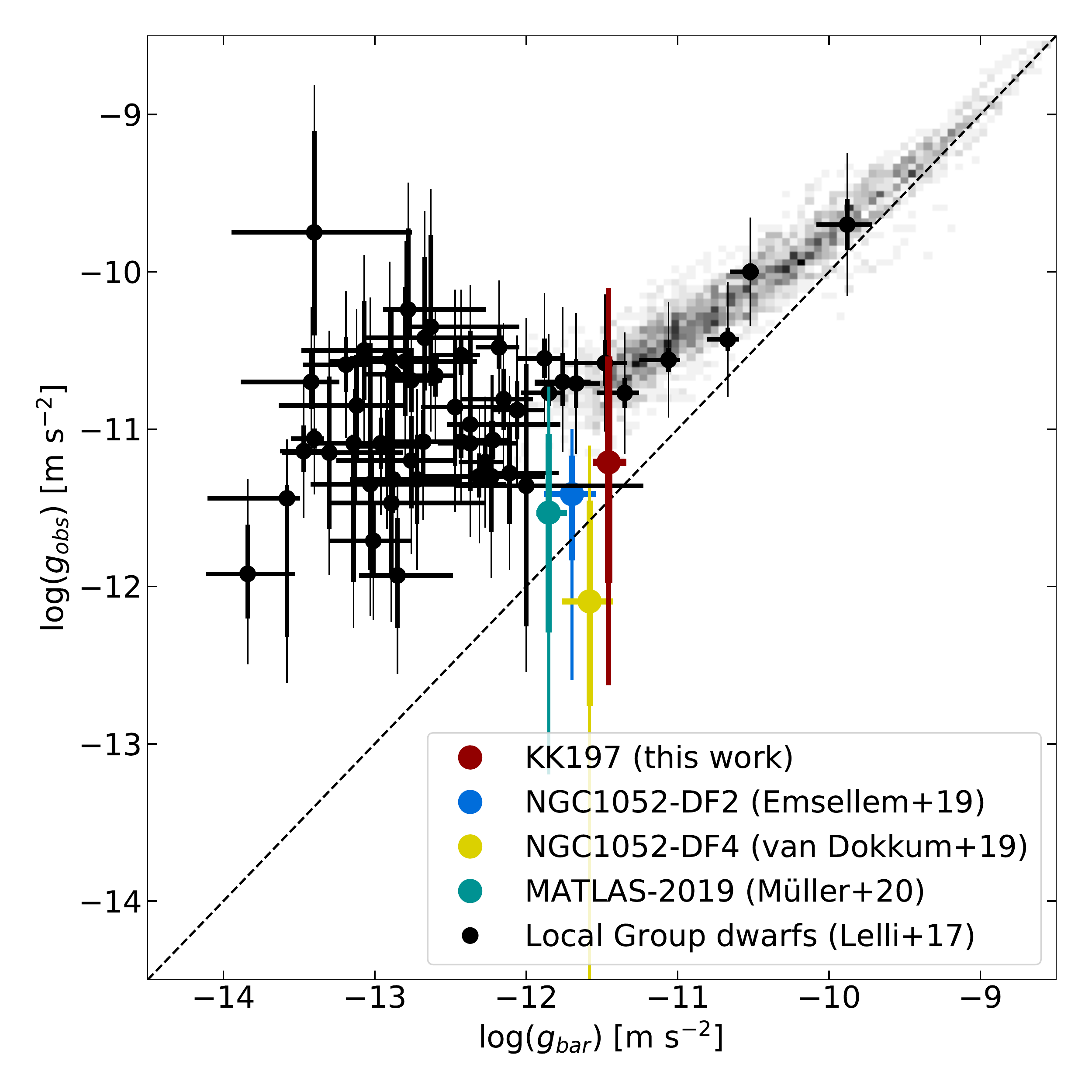}
    \caption{Radial acceleration relation of galaxies. The dashed line corresponds to unity, meaning that the baryonic acceleration (x-axis) is equal to the observed acceleration in the system (y-axis). {The gray scale represents $\sim$2700 spatially resolved measurements in rotationally supported galaxies \citep[spirals and dIrrs, see][]{2016PhRvL.117t1101M}}. The black dots correspond to pressure-supported dwarf galaxies in the Local Group as compiled in \citet{2017ApJ...836..152L} and come from different sources (e.g., \citealt{2007ApJ...670..313S,2008ApJ...675..201M,2009ApJ...690..453K,2009ApJ...704.1274W,2010ApJ...711..361G,2012ApJ...752...45T,2013ApJ...768..172C,2015ApJ...799L..13C,2014ApJ...793L..14M}). The cyan dot {represents} MATLAS-2019 \citep{2020A&A...640A.106M}. The blue and yellow lines are two {so-called ultra diffuse galaxies:} NGC1052-DF2 \citep{2019ApJ...874L..12D} and NGC1052-DF4 \citep{2019ApJ...874L...5V}. The red dot {shows our new measurements for} KK\,197. The thick and thin lines are the 1 and 2$\sigma$ uncertainties, respectively.}
    \label{fig:rar}
\end{figure}

{For spherical, pressure-supported systems, the observed acceleration can be computed as}
\begin{equation}
g_{\rm obs}=\frac{3\,\sigma_{\rm int}^2}{r_{1/2}},
\end{equation}
{and the baryonic gravitational field is given by}
\begin{equation}
g_{\rm bar}=\frac{\Gamma_V\,G\,L_V}{2\,r_{1/2}^2},
\end{equation}
where $\Gamma_V$ is the stellar mass-to-light ratio. {From the spectroscopy of KK\,197, we} measured $\Gamma_V=2.3^{+0.3}_{-0.3}$. {Fig.\,\ref{fig:rar} shows the location of KK\,197 on the RAR established using rotation-supported disk galaxies (gray color scale) together with measurements of other pressure-supported dwarf galaxies from the literature. KK\,197 follows the same RAR as rotation-supported galaxies within 2$\sigma$, independently of whether we include or exclude the PN in the estimate of the velocity dispersion.}

The dynamics of dwarf galaxies can be used to test alternative gravity models like modified Newtonian dynamics (MOND, \citealt{1983ApJ...270..365M}, see also its review by \citealt{2012LRR....15...10F}). In MOND, rather than adding dark matter to a galaxy to explain the high dynamical mass-to-light ratio, the law of gravity is modified such that the baryons produce the dark matter-like behavior. Due to the {nonlinear} modification of the Poisson equation in MOND, a so-called external field effect (EFE, see e.g., \citealt{2019MNRAS.487.2441H}) {emerges} when the galaxy is embedded in an external gravitational field. {If the internal acceleration is stronger than the external one, the galaxy can be treated as in isolation, however in the opposite case the EFE has to be considered and it will lower the expected internal acceleration. \citep{kroupa2018does,2018MNRAS.480..473F}}

\begin{figure}[ht]
    \centering
    \includegraphics[width=9cm]{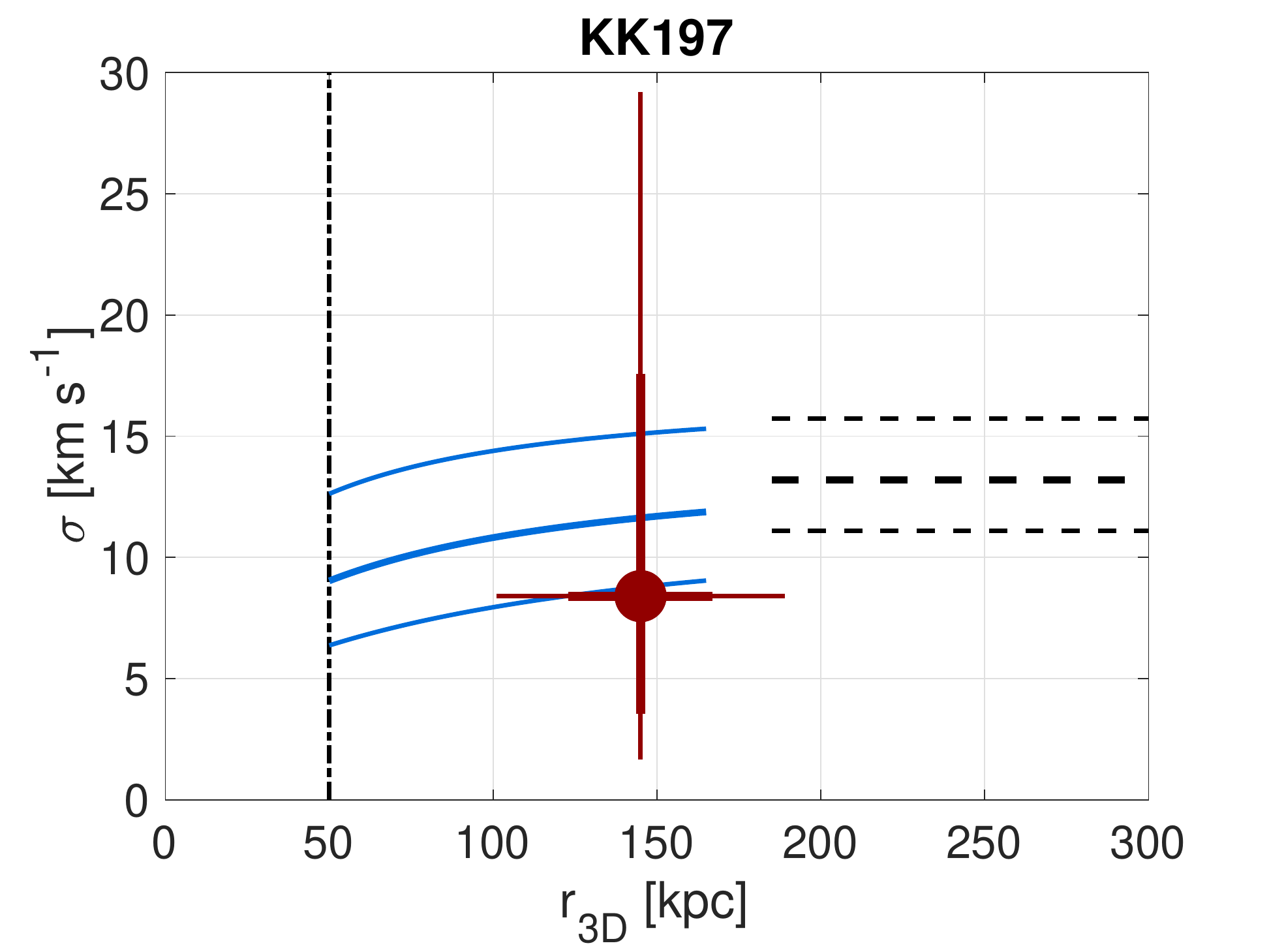}
    \caption{Predicted MOND velocity dispersion for KK\,197 as a function of its 3D separation {to Cen\,A}. The horizontal dashed lines show the MOND prediction (and its 1$\sigma$ uncertainty) for the isolated case, while blue lines correspond to the MOND prediction
    {considering} the EFE {from Cen\,A. The red dot {is the measured value considering the stellar clusters}}.     The 1$\sigma$ and 2$\sigma$ uncertainties are represented by the thick and thin red lines..
    }
    \label{fig:mond}
\end{figure}

Following the methodology presented in \citet{2019A&A...623A..36M}, we calculated the velocity dispersion for KK\,197 as a function of its 3D separation to Cen\,A. The measured separation is $r_{3D}=145\pm22$\,kpc \citep{2007AJ....133..504K,2013AJ....145..101K}.
In Fig.\,\ref{fig:mond}, we present the MOND prediction and compare it to the observations. In isolation, the expected velocity dispersion is $\sigma_{\rm MOND, isolated}=13.2^{+2.5}_{-2.1}$\,km s$^{-1}$. However, at this 3D separation {from Cen\,A}, the galaxy is already affected by the EFE, lowering the expected velocity dispersion to $\sigma_{\rm MOND,EFE}=11.5^{+4.5}_{-3.0}$\,km s$^{-1}$. 
{This prediction is in close agreement with the measured velocity dispersion of $8.5^{+9.3}_{-4.9}$ km s$^{-1}$ when the PN is excluded, as it is likely associated with Cen\,A. When the PN is included in the fit, the measured dispersion increases to $25.2^{+14.3}_{-7.6}$ km s$^{-1}$ (i.e., a 2$\sigma$ tension with the predicted MOND value).}\\

\subsection{H$\alpha$ emission of KK\,203}\label{Sec:KK203}

\begin{figure*}
    \centering
    \includegraphics[width=0.99\textwidth]{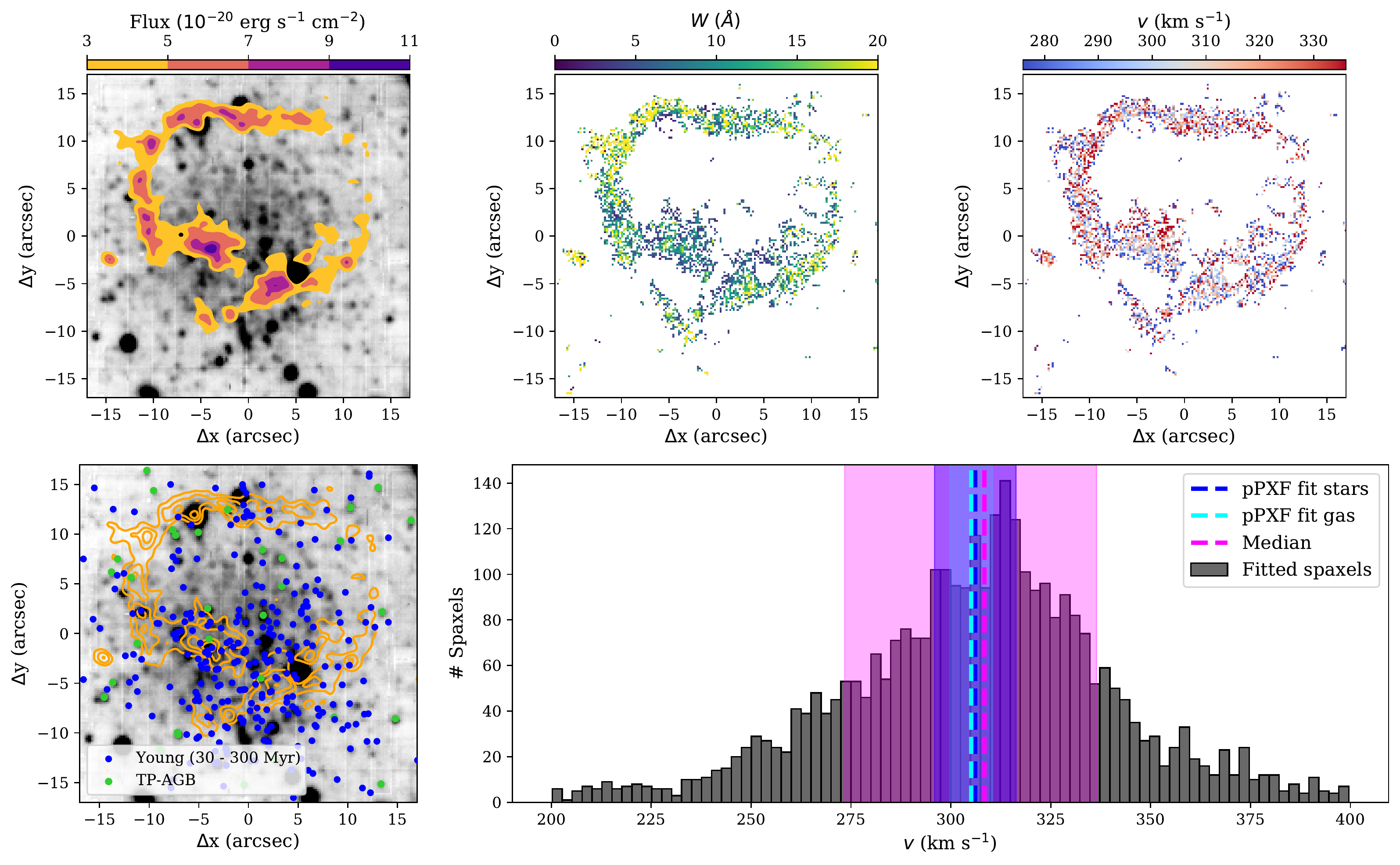}
    \caption{Dwarf galaxy KK\,203 with its H$\alpha$ {emission}.
    Top left: White-light image of KK\,203 with H$\alpha$ {intensity} contours. The contours refer to the residual H$\alpha$ emission, created by subtracting a continuum image from a narrow-band H$\alpha$ image obtained from the MUSE cube. Top-middle and top-right: equivalent width and line-of-sight velocity of the H$\alpha$ emission, obtained from fitting spaxels 
    with a Gaussian curve. Bottom-left: White light image with H$\alpha$ contour and young supergiants (blue) and TP-AGB stars overplotted (see Fig. \ref{fig:hst} for the selection). RGB stars are distributed randomly over the galaxy and are not shown here. Bottom-right: Histogram of line-of-sight velocities. The median with 1$\sigma$ uncertainties is highlighted in pink, light and dark blue refer to the results from the pPXF fit of the stellar and gas component to the integrated spectrum.} 
    \label{fig:ring}
\end{figure*}

\begin{figure}[ht]
    \centering
    \includegraphics[width=\linewidth]{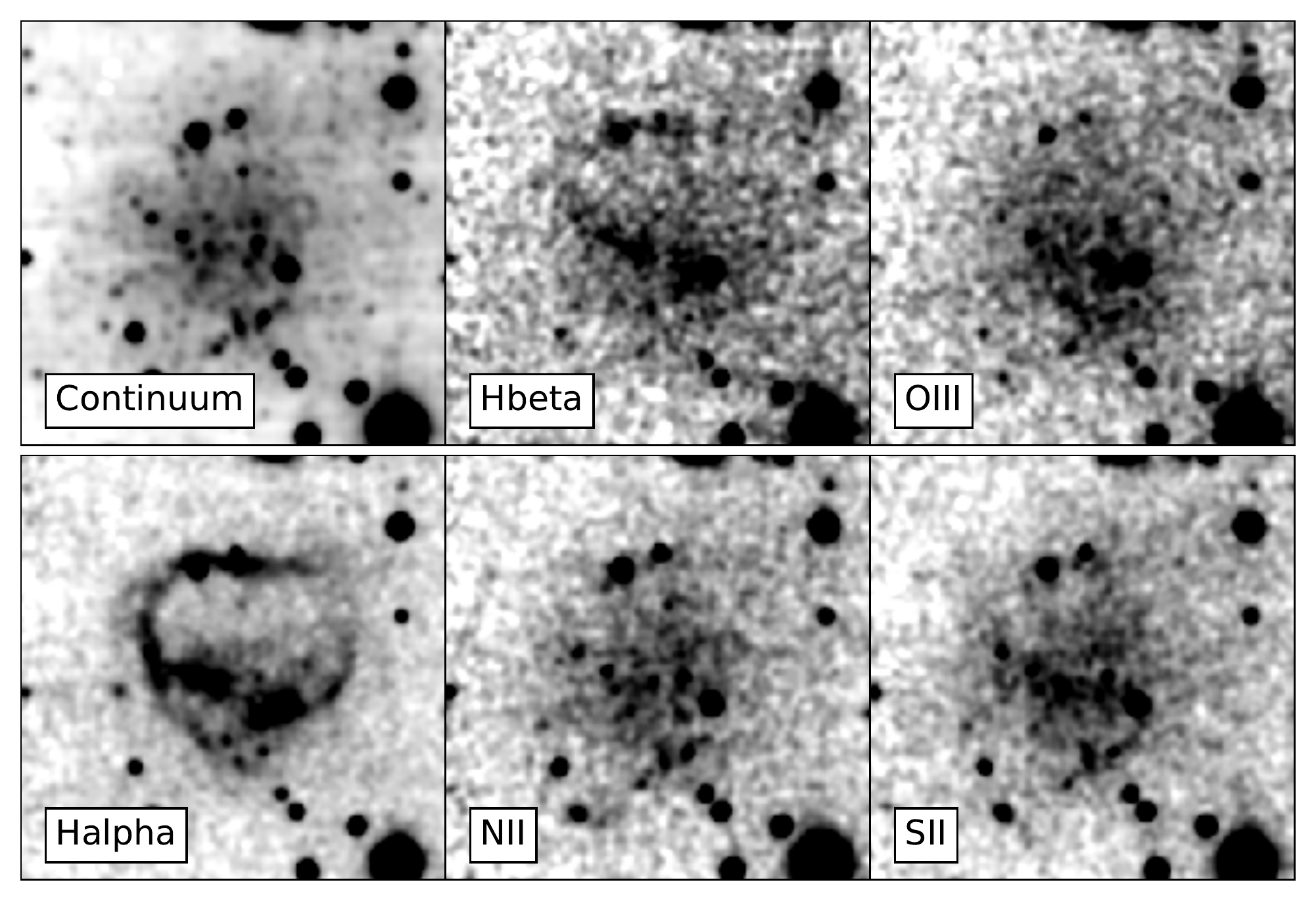}
    \caption{Dwarf galaxy KK\,203 and its emission regions. The images were produced as the sum of the MUSE cube for the continuum panel, and as single slices at the corresponding wavelengths for the emissions. To enhance the features, a small Gaussian convolution was applied.}
    \label{fig:emissions}
\end{figure}

\begin{figure*}
    \centering
    \includegraphics[width=0.48\textwidth]{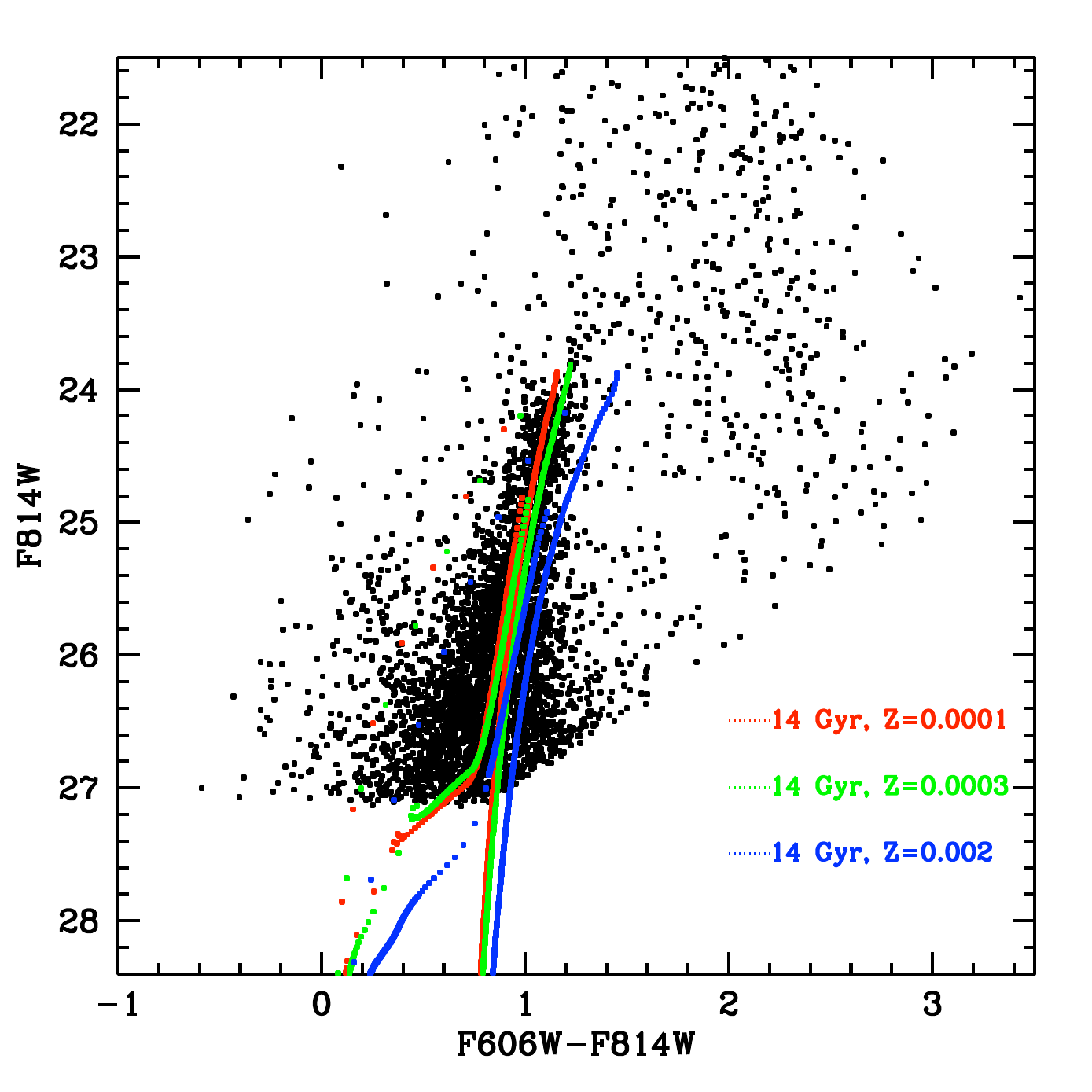}
    \includegraphics[width=0.48\textwidth]{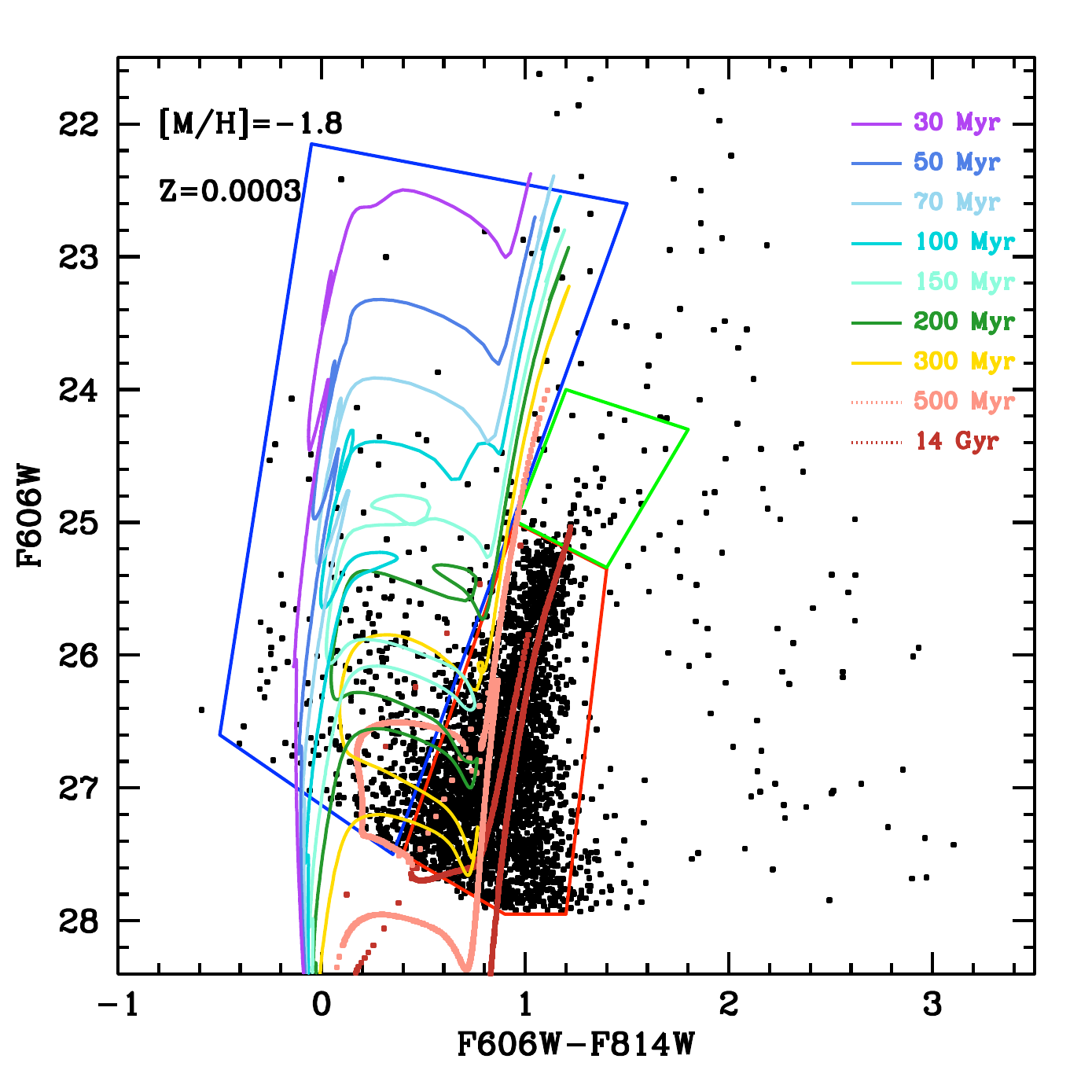}
    \caption{Left panel: $F814W$ versus $F606W-F814W$ color-magnitude diagram of KK~203 from the ACS HST imaging showing all sources identified as bona fide stars by the DOLPHOT PSF fitting routine. Overplotted are three BASTI isochrones for a 14 Gyr old population and metallicities as indicated in the legend. Right panel:  Statistically cleaned CMD, this time with $F606W$ on y-axis to emphasize the presence of young blue stars with ages ranging between 30-300 Myr (blue box). The areas of the CMD where the old RGB stars (red) and intermediate-age thermally pulsing asymptotic giant branch (TP-AGB) stars (green) are located are also indicated. A set of BASTI solar-scaled isochrones with the range of ages indicated in the legend does not include the TP-AGB phase. 
    }
    \label{fig:hst}
\end{figure*}

{Unexpectedly,} in the spectrum of one of the dSphs (KK\,203) we {detected} {conspicuous} H$\alpha$ emission, {forming}
an extended heart-shaped region {around}
the galaxy (see Fig.\,\ref{fig:ring}). {This} seems to be a continuous ring, somewhat off-centered from the galaxy, with a diameter of $\sim$24$''$ (440\,pc). On average, the H$\alpha$ velocities obtained from fitting individual spaxels are well in agreement with the systemic velocity of KK\,203,
{making their physical association extremely likely}. {Only H$\alpha$ and perhaps H$\beta$ emissions are detected forming this ring, other potential emission lines
%are consistent with 
{do not stand out above} the continuum of the galaxy (see Fig\,\ref{fig:emissions}).}

Where does this H$\alpha$ {emission} come from? 
{H$\alpha$ emission is typically associated with the presence of star formation, AGN, or shocks. In late-type spiral galaxies, on the order of 50\% of H$\alpha$ emission is coming from a warm diffuse component of the interstellar medium (ISM, \citealt{2007ApJ...661..801O}), which can be ionized by photons leaking from HII regions, evolved field stars, shocks, or cosmic rays. This so-called diffuse interstellar gas (DIG) or warm ionized medium is also observed in early-type galaxies \citep{1986AJ.....91.1062P,2004AJ....128.2758M,2014MNRAS.440.3491J}. The DIG has a lower electron density, higher electron temperature, and enhanced line ratios between collisionally excited and recombination lines when compared to HII regions. Unfortunately, the lack of unambiguous detection of  forbidden emission lines such as [SII] and [NII] within the same heart-shaped ring region means that we cannot use the usual diagnostics based on integrated spectra \citep{2010AJ....139..712L, 2019ARA&A..57..511K}.}

There is no detection of neutral hydrogen in HIPASS {at the location of KK\,203} \citep{2001MNRAS.322..486B}, so the {amount of cold gas (the fuel for any potential star formation)} is rather low {as expected for a dSph galaxy}. {However, the non-detection might also be due to the fact that the dwarf is smaller than the HIPASS spatial resolution element (15\,arcmin) and that the HIPASS detection limit is around M$_{\rm{HI}} \approx 10^7$\,M$_\odot$ {for dwarf galaxies in the Centaurus group \citep{2017A&A...597A...7M}}. } 

{There are no obvious individual HII regions detected in the MUSE data-cube. To ensure that we are not missing a possible low fraction of young or intermediate-age stars due to relatively low S/N of our spectroscopic data we carefully examined the available optical imaging from the HST archive. Based on DOLPHOT PSF fitting photometry \citep{DOLPHOT} applied to the ACS@HST images (HST program 13442, PI: Tully) we derive $F606W$ (wide $V$-band) versus $F606W-F814W$ (equivalent to $V-I$) color-magnitude diagram (CMD) shown in Fig.~\ref{fig:hst}. We first verified that the bulk of the stars in the CMD is well fit with an old and metal-poor population by overlaying 14 Gyr BASTI solar-scaled isochrones \citep{2004ApJ...612..168P} with metallicities that bracket the mean value derived from the MUSE spectrum. We find an excellent agreement between spectroscopic metallicity from the integrated MUSE spectrum and the photometric metallicity from the color of the red giant branch (RGB) sequence.

The foreground MW stars are primarily found at $F606W-F814W \ga 1.1$, which is redder than most of the KK~203 stars, but some compact blue galaxies may contaminate star counts at $F6060W-F814W \la 0.2$ and $F6060W \ga 25.5$, where young stars of KK\,203 are located. We cleaned the CMD from foreground/background stellar sources by applying statistical subtraction from a neighboring patch of the sky selected in the same image \citep[see][for more details]{MuellerTRGB2018}. The cleaned CMD still presents an excess of blue stars. The detection limit at $F814W \sim 27$ mag, visible as a diagonal limit in the $F606W$ versus $F606W-F814W$ CMD (right panel of Fig.~\ref{fig:hst}), implies that main sequence stars with ages older than $\sim 30$~Myr are not observable. The right panel of Fig.~\ref{fig:hst} shows the statistically cleaned CMD with the solar-scaled BASTI isochrones for [M/H]=$-1.8$\,dex and a range of ages. It shows that evolved supergiants younger than $\la 300$~Myr can be detected. Their short evolutionary lifetime leads to a relatively sparsely populated CMD area on the blue side of the RGB. We  verified by overplotting more metal-rich isochrones that there are likely no stars younger than $\sim 30$~Myr present in KK~203 and that stars indicated in {the blue selection box} span a range of ages between 30-300 Myr. Due to the age-metallicity degeneracy in the $VI$ photometric bands it is not possible to assign unambiguous ages to these blue stars. Deeper observations are needed to constrain the metallicity and thus derive more accurate ages for at least a handful of brightest supergiants. Such data could be obtained with adaptive optics assisted MUSE observations \citep[e.g.,][]{2020A&A...635A.134D}.

In Fig.~\ref{fig:hst} we identify regions of the CMD according to age: the RGB (red) contains predominantly old stars, the green area indicates the region where intermediate-age thermally pulsing asymptotic giant branch (TP-AGB) stars are located, and we show young supergiants in the blue selection box. After selecting stars on the statistically cleaned CMD of KK203 according to region, we plot them on top of the MUSE white-light cube image with color coding of dots reflecting the CMD region color {(except for RGB stars)}, and overplotting in cyan contours the H$\alpha$ gas (bottom left panel in Fig.\,\ref{fig:ring}). The youngest stars are most centrally concentrated and are displaced southwards with respect to the galaxy center.
Our conclusion from the analysis of the HST data is that, although there are some young stars with ages between 30-300 Myr, the low-level star formation is likely not sufficient to photoionize the extended H$\alpha$ emission in a ring surrounding KK~203.
 }

{Another common source of H$\alpha$ emission in massive galaxies is represented by supermassive black holes powering active galactic nuclei (AGN). In this case, however, we would expect the H$\alpha$ emission to be concentrated toward the galaxy center rather than forming an extended ring. Unfortunately, given the low metallicity of the galaxy, we do not detect oxygen and nitrogen emission lines, which could have been used to build BPT diagrams \citep{1981PASP...93....5B} and study the nature of the radiation field.}

Another possibility is that the H$\alpha$ emission comes from winds of AGB and post-AGB stars {or young supergiants}. For this purpose, we have used the HST photometry of this galaxy, plotted all potential AGB stars (extending the TRGB to brighter magnitudes) and post-AGB {and supergiant} stars (bluer than the RGB branch) over the collapsed MUSE cube, and compared it to the H$\alpha$ emission. The {candidate} AGB stars are roughly uniformly distributed over the galaxy, whereas the {candidate}
{bluer} stars are somewhat clustered toward the southwest, overlaying at least partially with the southern part of the H$\alpha$ ring. However, there is no clear correlation between these stars and the H$\alpha$ ring, especially for the northern part. If we {reverse} the approach and use the H$\alpha$ ring as a mask to select the stars in the CMD, mainly RGB stars are selected. This indicates that there is no evidence for a connection between AGB or post-AGB stars and the H$\alpha$ ring.

{Could the H$\alpha$ emission have an external or environmental origin?} One possibility is that {some interstellar gas in KK\,203} is excited by an exterior radiation field, namely that of Cen\,A. The 3D separation between Cen\,A and KK\,203 is $\approx$230\,kpc. 
This seems to be at a distance from Cen\,A  where the radiation density is too low for exciting/shocking the gas in the dwarf. Perhaps KK\,203 had a closer passage to Cen\,A in the past and {its interstellar medium was tidally shocked, so} what we see in H$\alpha$ is the aftereffect of this encounter. {A more speculative idea is that KK\,203 has accreted a much smaller dwarf galaxy or DM halo with some diffuse ionized gas, so the ring would represent a gaseous stream.}

{Yet another explanation could be} that we see an {old} supernova remnant (SNR). \citet{2020MNRAS.495.3592K} conducted an H$\alpha$ survey in dwarf galaxies with the 6-m BTA telescope of the Special Astrophysical Observatory. In their sample of 309 {irregular and transitional type} dwarf galaxies a number of different emission features were uncovered. 
Among them, {there} were so-called B-bubbles, which they describe as ring-like structures similar to SNRs. {About} 17\% percent of their dwarf galaxies contained such B-bubbles with an average  diameter of 120-240\,pc, which is somewhat smaller than what we observe here (we note though that \citet{2020MNRAS.495.3592K} simply assumed a distance of 5\,Mpc for all their targets and transformed the average angular size into a physical one, but also none of them is as large as the ring around KK\,203, private communication with I. Karachentsev). \citet{2019A&A...622A.129E} found two bubbles reminiscent of the one discovered here in the nearby dIrr Leo P (see their Fig.\,10). To confirm whether this heart-shaped {ring} in KK\,203 is an SNR, one would need to use the emission line ratios as diagnostics \citep{2019MNRAS.488..803M}, for which {our current} observations are too shallow. {Moreover, given the low metallicity of the system, it is unclear whether deeper observations could be able to detect other gas emission lines beyond the Balmer series of atomic hydrogen.} 
An SNR {is the most} satisfactory explanation {for the following reason.}
Assuming that the SNR expanded with a constant velocity of 100\,km s$^{-1}$ (\citealt{1976ApJ...208L..83K}), it would have needed $4\times10^6$ years to expand into its current form. {This is consistent with the age of some of the brightest blue and red supergiants observed in the HST CMD (right panel of Fig.~\ref{fig:hst}).}

\section{Summary and conclusions}

{We used MUSE spectroscopic observations to study the properties of a sample of dSphs in the Cen\,A galaxy group.}
Out of the 14 {targeted} objects, only two  were revealed to be background galaxies based on their redshifts of $\sim$0.01. 
The {remaining galaxies have}
been confirmed as Cen\,A members {based on their line-of-sight velocities. We found the following results:}
\begin{enumerate}
    \item {The integrated spectra of these 12 dSphs are consistent with old and metal-poor stellar populations,} as expected from observations of dSphs in the Local Group. {Moreover, the dSphs in the Cen\,A group follow a similar stellar metallicity-luminosity relation as dwarf galaxies in the Local Group.}
    \item  For the three brightest dSphs in our sample {(KK~197, KKs~55 and KKs~58)}, we found associated GCs. The specific frequency of GCs for these three dwarfs {is around 18}, which falls on the expected range from {previous} studies of other nearby galaxies.
    \item We found a PN near the dwarf galaxy KK\,197, {revealed by} its strong [OIII] and H$\alpha$ emission lines. {It is quite likely, however, that this PN belongs to the outer halo of Cen\,A based on dynamical considerations.}
    \item {For the brightest dSph -- KK\,197 -- we estimated the intrinsic velocity dispersion using discrete tracers. Depending on whether the PN is considered or not, the velocity dispersion ranges from 8.5 to 25.3 km s$^{-1}$ and the corresponding dynamical mass-to-light ratio from $\sim$4 to $\sim$37 M$_\odot$/L$_\odot$.}
    \item {KK\,197 lies on the same radial acceleration relation of rotationally supported galaxies within the errors. The measured velocity dispersion also agrees with the MOND prediction.}
    \item For one of our dSphs -- KK\,203 -- we found a {surprising} 
    H$\alpha$ emission, which forms a 400-pc wide,
    heart-shaped ring. {The H$\alpha$ emission is not powered by recent star-formation. Its origin remains unclear but several possibilities are discussed, including a central supermassive black hole, AGB and post-AGB stars, a $\sim 40$~Myr old supernova remnant, or past interaction/accretion events.} {Our favorite interpretation is that H$\alpha$ is powered by an SNR, but deeper data are necessary to establish that on a firmer ground.}
\end{enumerate}

{In a companion paper, we use the measured line-of-sight velocities of these 12 dSphs to study the overall dynamics of the Cen\,A group and, in particular, to confirm or disprove the presence of a rotating satellite system \citep{2018Sci...359..534M}.}

\begin{acknowledgements} 
{We thank the referee for the constructive report, which helped to clarify and improve the manuscript.}
O.M. is grateful to the Swiss National Science Foundation for financial support. The authors thank Aku Venhola, and Yvez Revaz for the interesting discussions on the H$\alpha$ content of KK\,203. This work has made use of BaSTI web tools. 
{M.S.P. and O.M. thank the Deutscher Akademischer Austauschdienst for PPP grant 57512596 funded by the Bundesministerium f\"ur Bildung und Forschung, and the Partenariat Hubert Curien (PHC) for PROCOPE project 44677UE. M.S.P. thanks the Klaus Tschira Stiftung and German Scholars Organization for support via a KT Boost Fund.} G.S.A acknowledges support for this work provided by NASA through grant number HST-SNAP-15922 from the Space Telescope Science Institute.
{Based on observations made with the NASA/ESA Hubble Space Telescope, obtained from the data archive at the Space Telescope Science Institute. STScI is operated by the Association of Universities for Research in Astronomy, Inc. under NASA contract NAS 5-26555.}
\end{acknowledgements}

\bibliographystyle{aa}
\bibliography{aanda}

\begin{appendix}
\section{Surface brightness photometry of KK\,203}
\label{photometry}
Due to the lack of surface brightness photometry of the dwarf galaxy KK\,203 in the literature, we have derived it here. 
We used archival $g$ and $r$ band imaging taken with the Dark Energy Camera. The stacked $g$ band image has an exposure time of 1000\,s, the $r$ band an exposure time of 60\,s. The data was processed by the standard DECam community pipeline \citep{2014ASPC..485..379V}. To model the surface brightness profile of the galaxy, we fit S\'ersic profiles to the galaxy using GALFIT \citep{2002AJ....124..266P}. We provided GALFIT with a segmentation map created by  MTObjects \citep{teeninga2013bi,teeninga2015improved} to mask foreground stars and background galaxies. The zero points were derived using The AAVSO Photometric All-Sky Survey (APASS) standard stars \citep{2009AAS...21440702H}. We derive an extinction corrected apparent magnitude of $m_g=16.28$\,mag with a color of $(g-r)_0=0.22$\,mag. The effective radius is $19.8$\,arcsec, which corresponds to 361\,pc at the distance of KK\,203 ($D=3.77$\,Mpc). The mean effective surface brightness is $\mu_{\rm eff,g}=24.75$\,mag. Using the color transformation by \citet{SloanConv}, we derive an absolute magnitude of $M_V=-11.7$ mag, which is slightly brighter than what is listed in the literature \citep{MuellerTRGB2019}.

\section{Properties of the background objects}\label{Sec:background}

Our MUSE data of KK\,198 and dw1315-45 revealed that these two galaxies are not dwarf galaxies associated with the Centaurus group but rather are background star forming galaxies.
Both KK\,198 and dw1315-45 show strong Balmer emission lines (H$\alpha$ and H$\beta$) as well as [NII], [SII], and [OIII] lines. KK\,198 has a redshift of 0.0128, dw1315-45 of 0.0100, which puts them at a distance of $\approx$50\,Mpc. The former is a face-on spiral galaxy, as already noted in the optical FORS2 images \citep{MuellerTRGB2019}, the latter must be an ultra-diffuse galaxy with an effective radius of $\approx2300$\,pc.

\section{Spectra of the targets}
\begin{figure*}[ht]
    \centering
    \includegraphics[width=\linewidth]{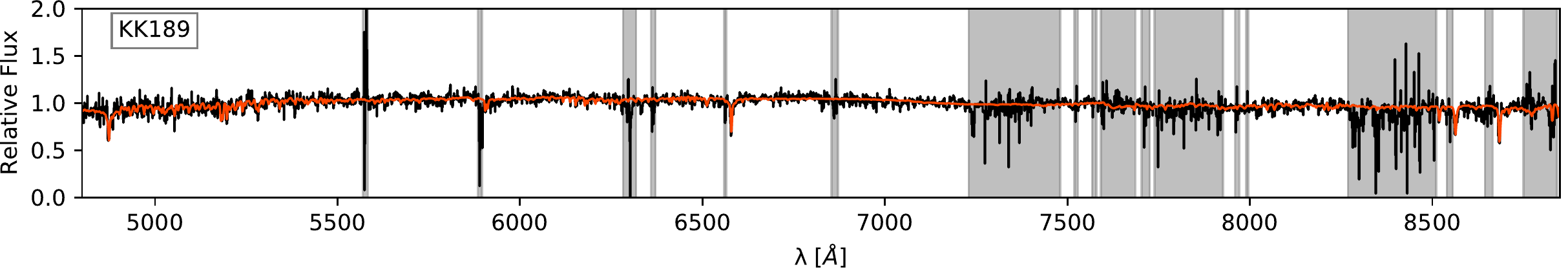}\\
    \includegraphics[width=\linewidth]{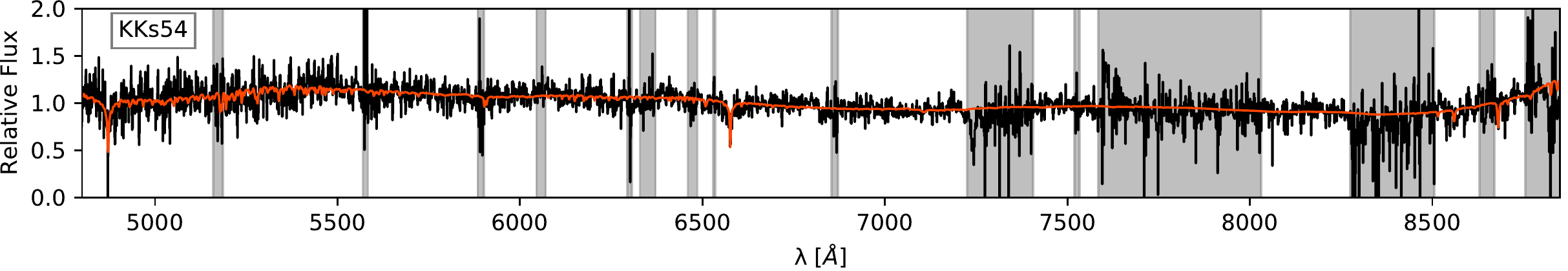}\\
    \includegraphics[width=\linewidth]{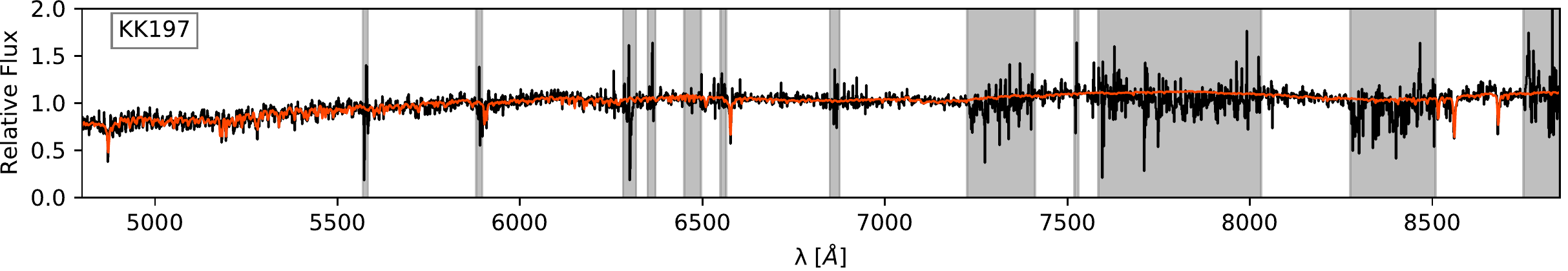}\\
        \includegraphics[width=\linewidth]{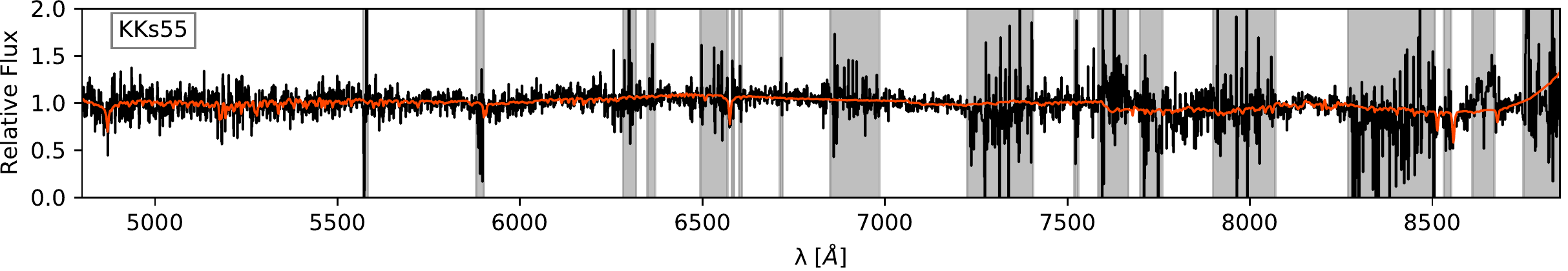}\\
    \includegraphics[width=\linewidth]{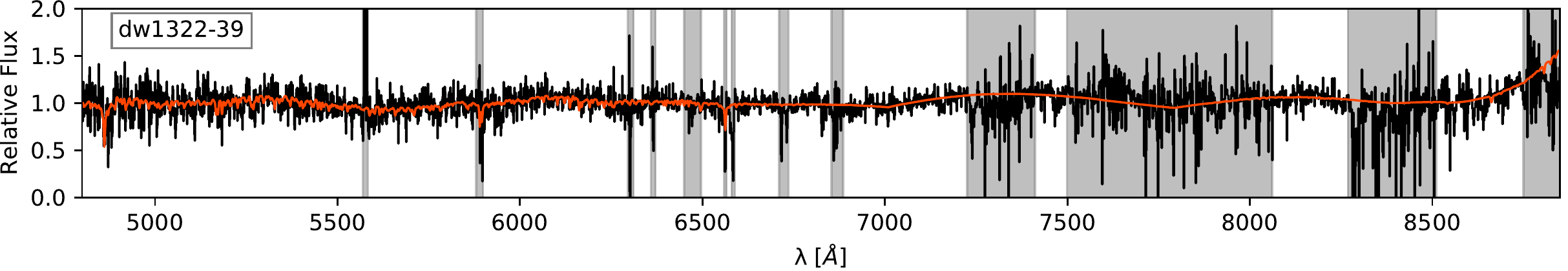}\\
    \includegraphics[width=\linewidth]{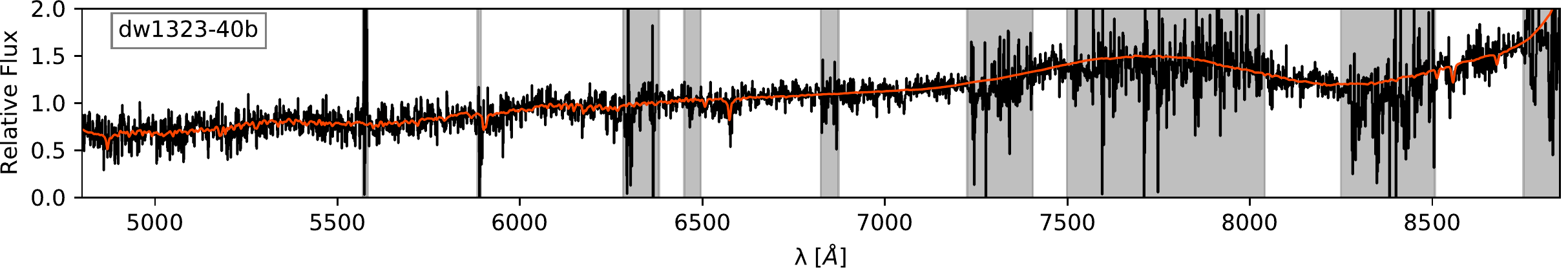}\\
    \includegraphics[width=\linewidth]{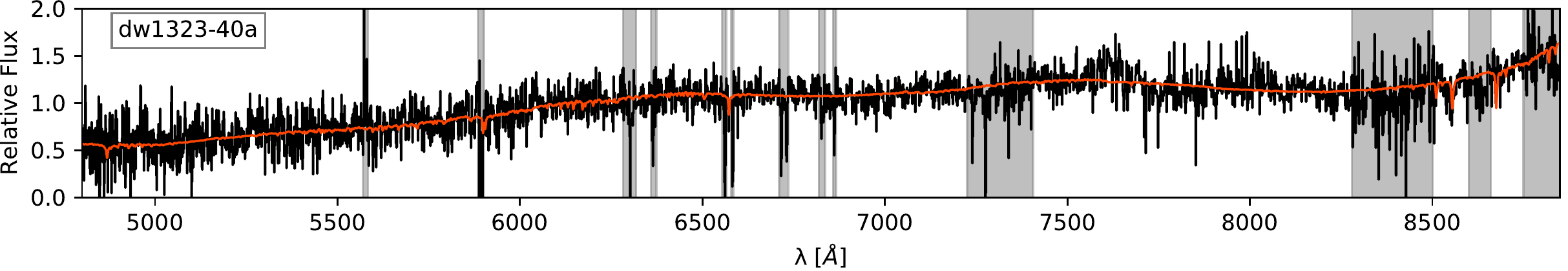}\\
    \caption{{Integrated MUSE} spectra (black) of all the observed targets. The gray area are masked regions, the red lines correspond to the best-fit from pPXF.}
    \label{fig:spectra_all1}
\end{figure*}

\begin{figure*}[ht]
    \centering
    \includegraphics[width=\linewidth]{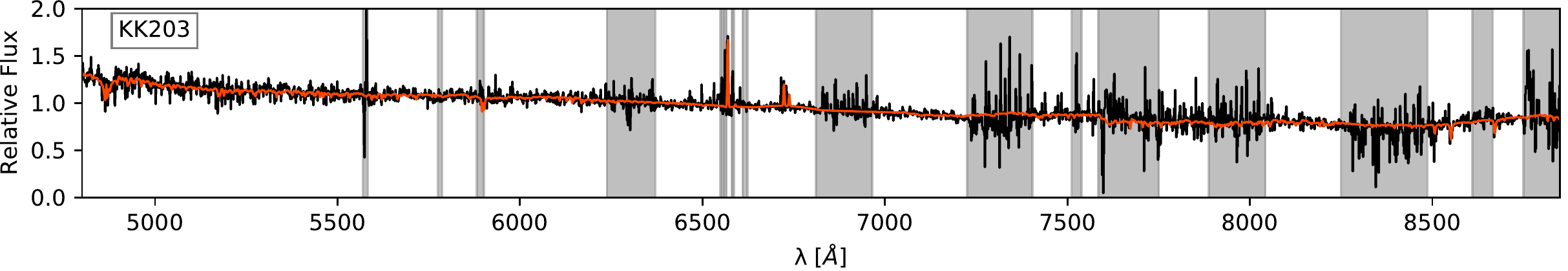}\\
    \includegraphics[width=\linewidth]{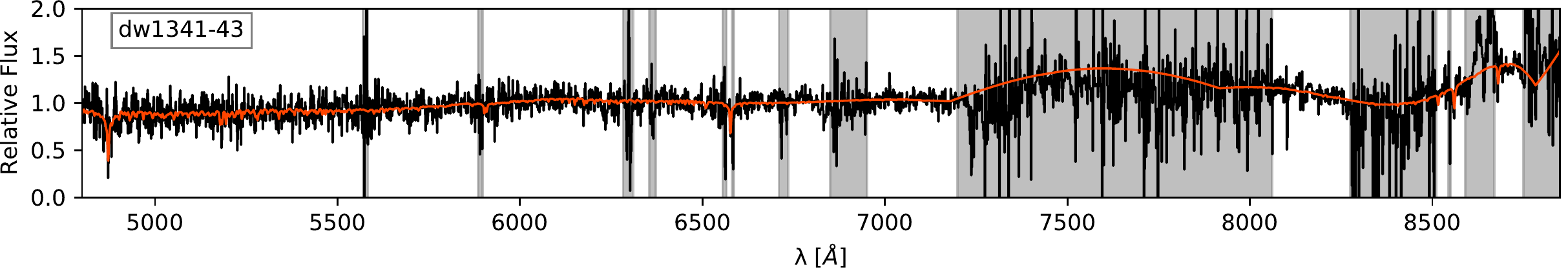}\\
    \includegraphics[width=\linewidth]{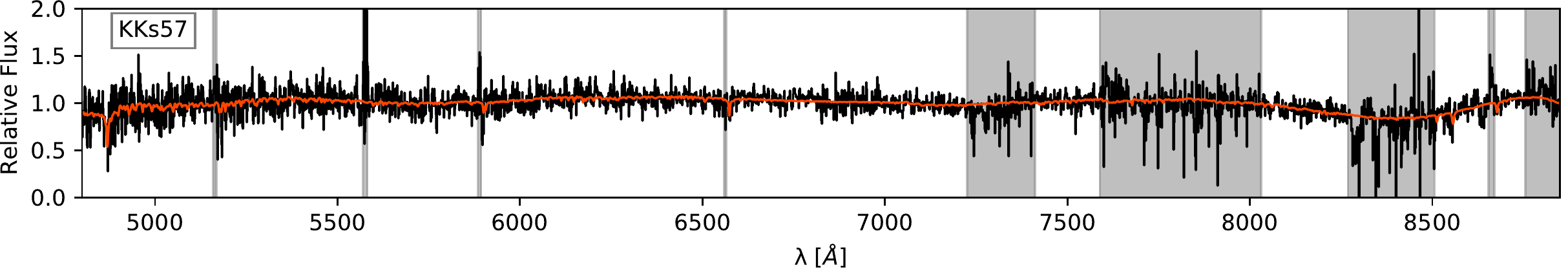}\\
    \includegraphics[width=\linewidth]{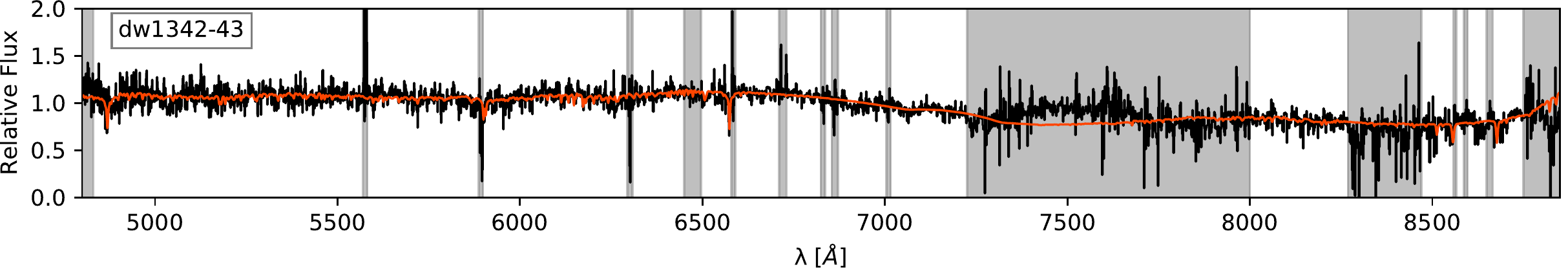}\\
    \includegraphics[width=\linewidth]{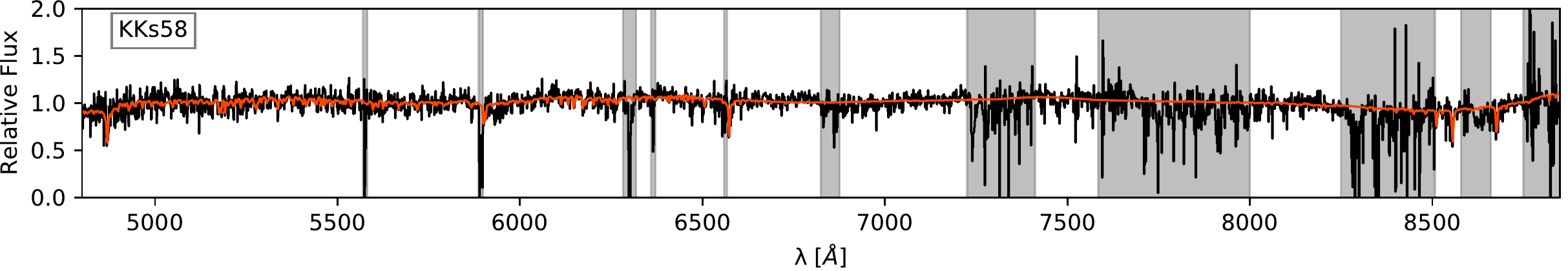}\\
    \includegraphics[width=\linewidth]{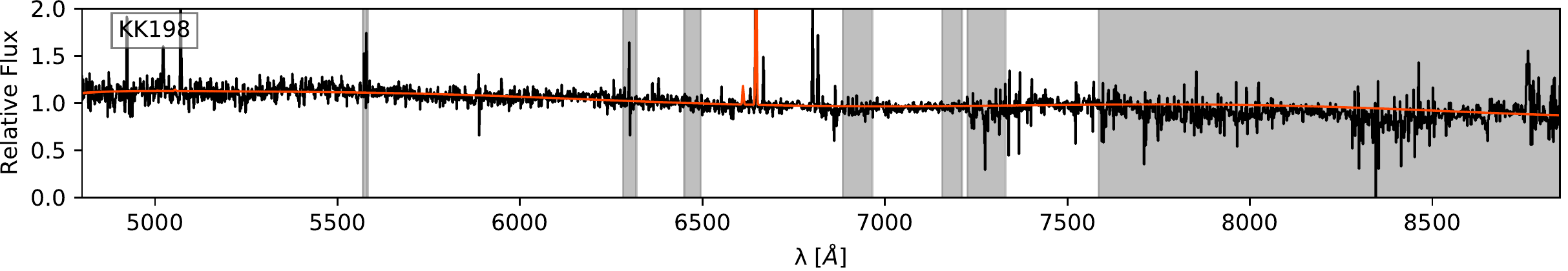}\\
            \includegraphics[width=\linewidth]{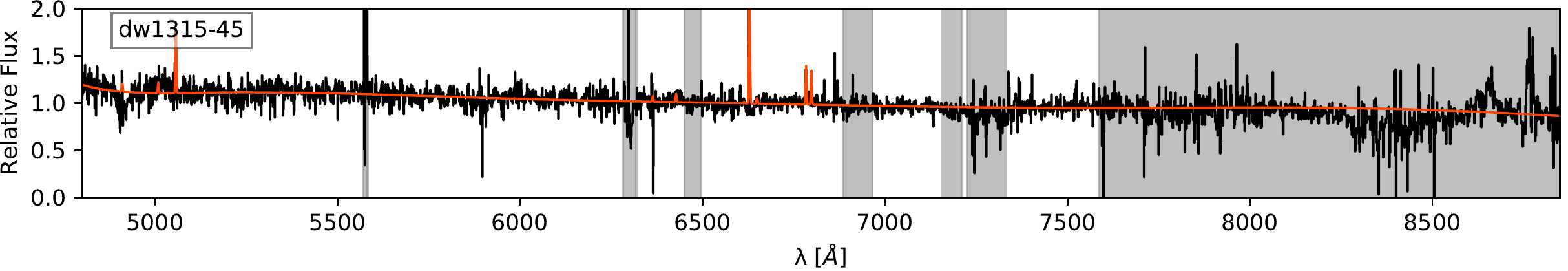}\\
    \caption{{Integrated MUSE} spectra (black) of all the observed targets. The gray area are masked regions, the red lines correspond to the best-fit from pPXF.}
    \label{fig:spectra_all2}
\end{figure*}

In Fig\,\ref{fig:spectra_all1} and  Fig\,\ref{fig:spectra_all2} we present all the 
{integrated} spectra of our targets. The best fit provided by pPXF includes the stellar absorption and emission lines. {Two targets (KK\,198 and dw1315-45) are clearly background star-forming galaxies with strong emission lines.}

\end{appendix}

\end{document}